\documentclass[11pt, letterpaper]{article}
\usepackage[letterpaper, margin=1in]{geometry}

\usepackage{cite}
\usepackage{amsmath,amssymb,amsfonts, amsthm}
\usepackage{algorithmic}
\usepackage{threeparttable}
\usepackage{graphicx}
\usepackage{textcomp}
\usepackage{bbm}
\usepackage{enumitem}
\usepackage{dsfont}
\usepackage[font=small, labelfont=bf]{caption}
\usepackage{parskip}
\usepackage[table]{xcolor}
\usepackage{authblk}
\usepackage[most]{tcolorbox}

\usepackage{booktabs}
\usepackage{multirow}
\usepackage{adjustbox}
\usepackage{tabularx}
\usepackage{hyperref}
\definecolor{darkblue}{rgb}{0,0.08,0.45}

\hypersetup{
    colorlinks=true,
    linkcolor=darkblue,  
    citecolor=darkblue, 
    urlcolor=darkblue,  
    pdfborder={0 0 0}
}

\usepackage{etoolbox}
\AtBeginEnvironment{table}{\small}

\newtheorem{theorem}{Theorem}
\newtheorem{proposition}[theorem]{Proposition}

\newtheorem{corollary}{Corollary}
\newtheorem{definition}{Definition}

\newtcolorbox{leftbar}[1]{
    blanker,
    borderline west={0.7pt}{0pt}{black},
    left=6pt,
    top=4pt,
    bottom=4pt,
    before skip=8pt,
    after skip=8pt,
    fontupper=,
    title={\textbf{Running Example: #1}},
    coltitle=black,
}

\begin{document}

\title{Reliable Estimation of Differential Privacy under Selection Bias}

\date{}

\author[1]{Valentin Dorseuil\thanks{Corresponding author: \texttt{valentin.dorseuil@ens.psl.eu}}\thanks{Work done while visiting the Vector Institute.}}
\author[2]{Meenatchi Sundaram Muthu Selva Annamalai}
\author[3]{Jamal Atif}
\author[1]{Olivier Cappé}
\author[4,5]{Nicolas Papernot}
\affil[1]{DI ENS, École Normale Supérieure, Université PSL, CNRS}
\affil[2]{University College London}
\affil[3]{CMAP, École Polytechnique, Institut Polytechnique de Paris}
\affil[4]{Vector Institute, University of Toronto}
\affil[5]{INRIA}
\renewcommand\Authands{ and }
\renewcommand\Affilfont{\small}

\maketitle

\begin{abstract}
    Statistical privacy audits are essential for verifying the theoretical privacy guarantees of differentially private (DP) mechanisms. Auditors typically use Membership Inference Attacks (MIAs) to estimate a lower bound on the true privacy leakage of DP mechanisms at a specified confidence level, which should not exceed the theoretical privacy upper bound guaranteed by DP.
    However, we found that many auditing implementations in prior work introduce a selection bias when running MIAs, which leads to statistically invalid and overestimated lower bounds. 
    Because most audits are not tight, this bias has thus far gone unnoticed: the estimated lower bounds remained well below the theoretical upper bounds.
    Nevertheless, accurately evaluating this lower bound is critical, as a flawed statistical assessment could lead to false alarms on correct mechanisms or an incorrect evaluation of the guarantee's tightness.

    This work employs tools from the statistical literature to correct this bias. We present an overview of the different correction methods and assess their respective advantages. Furthermore, we perform a comprehensive re-analysis of results returned by prior audit implementations and show that without statistical correction, these audits violate the theoretical upper bounds up to four times more often than their claimed confidence level implies. With correction applied, the resulting lower bounds are statistically valid, and violations occur at no more than the nominal rate.
\end{abstract}

\normalsize
\section{Introduction}

Machine learning models, and particularly deep neural networks, are known to leak private information, e.g., by memorizing their training dataset~\cite{yeomPrivacyRiskMachine2018, carliniSecretSharerEvaluating2019, feldmanWhatNeuralNetworks2020}. When models are trained on sensitive data, this memorization can pose severe privacy concerns, as an adversary with access to the model may be able to recover information about individual training points~\cite{carliniExtractingTrainingData2021, shokriMembershipInferenceAttacks2017}, infer sensitive attributes~\cite{modelInversion2015}, or even reconstruct training examples themselves~\cite{balle2022reconstructing, carliniExtractingTrainingData2021}.

To address this issue, Differential Privacy (DP)~\cite{dworkDifferentialPrivacy} has become the standard framework for training models with formal privacy guarantees. It guarantees, up to a certain level, that a third-party attacker cannot distinguish whether any single data point was used in the training dataset or not. To train deep learning models under this guarantee, practitioners commonly use Differentially Private Stochastic Gradient Descent (DP-SGD)~\cite{abadiDeepLearningDifferential2016}. It modifies the standard SGD algorithm by clipping per-sample gradients, to bound the influence of any single training point, and injecting Gaussian noise at each iteration. It comes with a privacy accountant~\cite{abadiDeepLearningDifferential2016, wangSubsampledRenyiDP2019, koskelaComputingTightDifferential2019, zhuOptimalAccounting2022, doroshenkoConnectDotsTighter2022} that computes the level of noise that needs to be added to these gradients to obtain a target privacy guarantee.

However, a privacy accountant's guarantee only holds if the implementation matches its mathematical assumptions. In practice, this is not always the case: incorrect noise calibration, faulty subsampling, or unbounded sensitivity in the accountant design, can all lead to violations of the claimed privacy parameters~\cite{jagielskiAuditingDifferentiallyPrivate2020a, ceberePrivacyTheoryBugs2026, tramerDebuggingDifferentialPrivacy2022}.
Privacy auditing addresses this gap: rather than trusting the claimed upper bound, an auditor empirically estimates a \emph{lower bound} on the true privacy parameter, and checks that it does not exceed the claimed one. In a regulatory context, it also serves an additional purpose. It can verify that a practitioner has not, whether deliberately or mistakenly, deployed a weaker privacy guarantee than the one mandated by a regulator or a data protection agreement. In this scenario, providing a statistically reliable audit is important. A flawed statistical procedure could either miss a genuine violation or falsely flag a compliant implementation.

The standard tool for estimating this lower bound is the Membership Inference Attack (MIA)~\cite{shokriMembershipInferenceAttacks2017}, which attempts to determine whether a specific record was part of the training set or not. An MIA assigns each record a score, and predicts its membership by thresholding this score. As the DP framework bounds the error rates of any possible MIA, the auditor uses the empirical error rates to  estimate a lower bound on the privacy parameter. 
Framed this way, privacy auditing is essentially a parametric statistical estimation problem: the auditor observes a finite sample of attack outcomes, drawn from an unknown underlying parametric distribution, and uses it to estimate a lower bound on the distribution's parameter, with a specified confidence level.

However, our analysis found that confidence intervals were improperly used in multiple prior audits. A common practice in existing auditing implementations is to evaluate an MIA at \emph{several} candidate thresholds and report the lower bound obtained from whichever threshold performs best. While it might seem intuitive, as a stronger attack should give a tighter bound, it is statistically invalid without appropriate correction on the confidence level. Choosing the threshold after observing which attack performs the best introduces a data-dependent selection bias, which incorrectly inflates the claimed lower bound, a problem that post-selection inference is designed to address~\cite{validPostSelection}.

To correct this, we adapt tools from the multiple-comparisons literature, ensuring that reported lower bounds hold at the claimed confidence level. We emphasize that our contribution is orthogonal to works on tightening empirical audits: we do not propose a stronger attack, and our correction applies on top of any existing MIA-to-lower-bound pipeline.
Our focus is instead the statistical validity of the derived bound. This matters because an inflated lower bound is not only imprecise, it can mistakenly claim that a DP guarantee has been violated. We also stress that we do not derive new statistical results. We apply established multiple-comparisons tools to a setting where they had not previously been used.

The primary contributions of this paper are summarized as
follows:

\begin{enumerate}[label=\roman*)]
\item We formalize the auditing framework for differential privacy as a statistical estimation problem, and provide multiple valid methods to compute lower bounds on privacy parameters at a given confidence level;
\item We empirically evaluate both our proposed methods and the invalid ones commonly used in practice, under known theoretical distributions, to quantify the bias induced by selecting the maximizing threshold; 
\item We recompute the bounds of previous auditing papers with the newly proposed valid methods and compare them with the previously reported invalid bounds. We find that without correction, these audits violate the theoretical upper bounds up to four times more often than their claimed confidence level implies. Valid methods yield lower bounds that are $5\%$ to $93\%$ smaller than the previously reported ones.
\end{enumerate}

\section{Related Work}
The following section reviews previous research related to our topic.

\subsection{Privacy Auditing of DP Mechanisms}
Auditing differentially private mechanisms was first introduced with statistical tests to detect violations of DP in~\cite{ding2018detecting}. It was brought to machine learning in~\cite{jagielskiAuditingDifferentiallyPrivate2020a}, where the authors use MIAs to derive an empirical lower bound on the mechanism's privacy parameter.
Since then, several works have improved the quality of audits, relying on more powerful attacks, such as calibrated-thresholds attacks~\cite{yeEnhancedMembershipInference2022}, LiRA~\cite{carliniMembershipInferenceAttacks2022}, RMIA~\cite{zarifzadehLowCostHighPowerMembership2024} or quantile-based MIA~\cite{liuEnhancingOnerunPrivacy2025}. Another line of work focuses on how MIA results are converted into an estimate of the lower bound, using Bayesian methods~\cite{zanella-beguelinBayesianEstimationDifferential2022} or $\mu$-GDP~\cite{dongGaussianDifferentialPrivacy2019} to obtain tighter bound estimates~\cite{nasrTightAuditingDifferentially2023}. Generally, audits have also proven useful for discovering bugs in DP implementations~\cite{tramerDebuggingDifferentialPrivacy2022, nasrTightAuditingDifferentially2023}.

Other works tried to improve the tightness of the audits by targeting specific threat models. These include increasingly powerful adversaries~\cite{nasrAdversaryInstantiationLower2021}, gradient injections in the white-box setting~\cite{jagielskiAuditingDifferentiallyPrivate2020a} and in the hidden state setting~\cite{cebereTighterPrivacyAuditing2025}, and adversarial initialization of the weights in the black-box setting~\cite{annamalaiNearlyTightBlackBox2024}. 
A further line of work performs audits in a single training run, by randomizing the inclusion of the audited points, the ``canaries''. Introduced in~\cite{steinkePrivacyAuditingOne2023}, this approach was extended to the $f$-DP framework~\cite{dongGaussianDifferentialPrivacy2019} in~\cite{mahloujifarAuditingfDifferentialPrivacy2024}, adapted to federated learning in~\cite{andrewOneshotEmpiricalPrivacy2024}, and more recently improved by analyzing the white-box case~\cite{agrawalLetsAskGauss2026}. 
Going further, zero-run approaches audit a model without any additional training, either by relying on holdout data that were never seen during training~\cite{kazmiPANORAMIAPrivacyAuditing2024} or by correcting for the temporal distribution shift between members and non-members~\cite{ceberePrivacyAuditingZero2026}.

\subsection{Threshold Selection in Prior Audits}
Most audits based on MIAs follow the same pipeline: the auditor trains models with and without the canaries, computes an attack score for each, and turns the scores into guesses using a decision threshold $\eta$. The resulting empirical error rates are then converted into a lower bound on the privacy parameter. This bound is valid at the nominal confidence level if $\eta$ is fixed independently of the samples used to build the confidence intervals. In practice, however, $\eta$ is often chosen after observing the audit outcome without any correction on the confidence interval.
Several works appear to report the threshold yielding the best lower bound, without applying any corrections~\cite{nasrTightAuditingDifferentially2023, annamalaiNearlyTightBlackBox2024, cebereTighterPrivacyAuditing2025, maddockCANIFECraftingCanaries2023, zanella-beguelinBayesianEstimationDifferential2022}.
One-run audits face an analogous issue: they fix a number of guesses rather than a threshold, and choose the number that maximizes the lower bound~\cite{steinkePrivacyAuditingOne2023, mahloujifarAuditingfDifferentialPrivacy2024}.

A few works avoid this issue by selecting the threshold on a separate split of the data~\cite{jagielskiAuditingDifferentiallyPrivate2020a, askinGeneralPurposefDPEstimation2025, ganevTightAuditingDifferential2026, changEfficientPrivacyAuditing2024, yeEnhancedMembershipInference2022, luGeneralFrameworkAuditing2023}. Others do not clearly state how the threshold is chosen~\cite{nasrAdversaryInstantiationLower2021, ceberePrivacyAuditingZero2026, annamalaiItsOurLoss2024a}. Finally, some works are not affected because they analyze the audit without selecting a threshold~\cite{koskelaAuditingDifferentialPrivacy2025, agrawalLetsAskGauss2026}.

\subsection{Multiple Comparisons and Post-Selection Inference}

The statistical literature has long recognized that inference is invalidated when the reported quantity is selected from many candidates using the same data. A confidence interval only has its nominal coverage for a parameter fixed in advance. If the parameter is chosen after looking at the data, for instance by reporting the best of many comparisons, the coverage of the reported interval can fall well below the nominal level. This was studied early in the context of multiple comparisons~\cite{scheffeMethod, bonferroniCorrection}.

The easiest approach is to control the \emph{family-wise} error rate, i.e., to make all the intervals simultaneously valid. The simplest solution is the Bonferroni correction~\cite{bonferroniCorrection}, which follows from the union bound. It requires no assumption on the dependence between the intervals but can be conservative. Other approaches control the false discovery rate~\cite{benjaminiControlling1995}, which limits the expected proportion of false rejections rather than the coverage, or the false coverage rate~\cite{benjaminiFalse2005}, which limits the expected proportion of selected intervals that fail to cover their parameter.

Post-selection inference generalizes this idea to arbitrary data-driven selection. Some methods build confidence intervals that are valid simultaneously over all possible selections~\cite{validPostSelection}, while others condition on the selection event~\cite{lee2016exact, fithian2014optimal}. However, the simplest valid approach remains data splitting~\cite{CoxDataSplitting1975}. It selects on one part of the data and computes the interval on the other, requiring no assumptions but wasting samples.

\section{Auditing Framework}
In this section, we show how auditing differential privacy can be framed as a statistical estimation problem: a DP guarantee constrains what any hypothesis test can achieve against a mechanism, and an audit turns the scores of an attack into a confidence lower bound on the privacy parameter. 
We also introduce a running example used to illustrate each step of the methodology.

\subsection{Differential Privacy}

To start, we recall the standard definition of Differential Privacy (DP)~\cite{dworkDifferentialPrivacy}. Consider a mechanism $\mathcal{M}: \mathcal{D} \to \mathcal{W}$, where $\mathcal{W}$ denotes the output space.
\begin{definition}[Differential Privacy~\cite{dworkDifferentialPrivacy}]

$\mathcal{M}$ satisfies $(\varepsilon, \delta)$-DP if, for all adjacent datasets $D_0, D_1 \in \mathcal{D}$, i.e., differing by the addition or removal of a single record, and all measurable sets $S \subseteq \mathcal{W}$,
\begin{equation}
    \mathbb{P}\left(\mathcal{M}(D_0) \in S\right) \le e^{\varepsilon} \mathbb{P}\left(\mathcal{M}(D_1) \in S\right) + \delta.
\end{equation}
\end{definition}
When $\delta = 0$, we recover pure $\varepsilon$-DP; $\delta > 0$ gives approximate DP. These notions, and the GDP notion below, are variants of the same idea: each framework limits how well any test can distinguish $\mathcal{M}(D_0)$ from $\mathcal{M}(D_1)$, and they differ only in how that limit is parameterized.

Gaussian Differential Privacy ($\mu$-GDP)~\cite{dongGaussianDifferentialPrivacy2019} is an alternative framework that characterizes privacy directly via hypothesis testing. For any statistical test $\phi$, attempting to distinguish $\mathcal{M}(D_0)$ from $\mathcal{M}(D_1)$, it bounds the trade-off between the false positive rate $\alpha_\phi$ and false negative rate $\beta_\phi$.

\begin{definition}[Trade-off Function~\cite{dongGaussianDifferentialPrivacy2019}]
\label{def:tradeoff}
For any two probability distributions $P$ and $Q$ on the same space, define the trade-off function $T (P, Q) : [0, 1] \to [0, 1]$ as  $$T (P, Q)(\alpha) = \inf_{\phi} \{\beta_\phi : \alpha_\phi \leq \alpha\},$$  where the infimum is taken over all (measurable) rejection rules.
\end{definition}

\begin{definition}[$\mu$-Gaussian Differential Privacy~\cite{dongGaussianDifferentialPrivacy2019}]
\label{def:mu-gdp}
A mechanism $\mathcal{M}$ satisfies $\mu$-GDP, for $\mu \ge 0$, if for all adjacent datasets $D_0, D_1 \in \mathcal{D}$,
\begin{equation}
    T(\mathcal{M}(D_0), \mathcal{M}(D_1)) \ge G_\mu,
\end{equation}
where $G_\mu = T(\mathcal{N}(0,1), \mathcal{N}(\mu,1))$ is the trade-off function between two unit-variance Gaussians with means $0$ and $\mu$.
\end{definition}

$\mu$-GDP is particularly convenient for auditing, as it summarizes the privacy guarantee by a single parameter $\mu$ and approximates the trade-off curves of DP-SGD~\cite{gomezGaussianDPReporting2026} well. It also admits a tight, closed-form conversion to $(\varepsilon,\delta)$-DP. This matters because $(\varepsilon,\delta)$-DP is the most widely used notion in practice.

\begin{corollary}[$\mu$-GDP to $(\varepsilon, \delta)$-DP conversion~\cite{dongGaussianDifferentialPrivacy2019}] A mechanism is $\mu$-GDP iff it is $(\varepsilon, \delta(\varepsilon))$-DP for all $\varepsilon\geq 0$, where:
\begin{equation}
\label{eq:eps_to_gdp}
    \delta(\varepsilon) = \Phi \left( -\frac{\varepsilon}{\mu} + \frac{\mu}{2} \right) - e^\varepsilon \Phi \left(-\frac{\varepsilon}{\mu} - \frac{\mu}{2}\right)
\end{equation}
\end{corollary}
Finally, both frameworks are preserved by \emph{post-processing}: a transformation of the output of a mechanism is at least as private as the mechanism itself.

\subsection{Statistical Estimation}
\label{subsec:stat_estimation}
We recall the statistical notions used throughout the paper. Let $\tau \in \mathbb{R}$ be a fixed, unknown parameter, and let $X = (X_1, \dots, X_n)$ be random observations drawn from a distribution that depends on $\tau$. The goal is to bound $\tau$ using only the observations $X$.

\begin{definition}[Confidence bound]
\label{def:conf_bound}
Let $\gamma \in (0, 1)$. A statistic $\underline{\tau} = \underline{\tau}(X)$ is a \emph{lower bound} on $\tau$ with \emph{confidence level} $1-\gamma$, if for every possible value of $\tau$,
\begin{equation}
\label{eq:coverage}
    \mathbb{P}(\tau \ge \underline{\tau}) \ge 1-\gamma.
\end{equation}
An \emph{upper bound} $\overline{\tau}$, with confidence $1-\gamma$, is defined similarly, with $\mathbb{P}(\tau \le \overline{\tau}) \ge 1-\gamma$.
\end{definition}

Here, $\tau$ is fixed and $\underline{\tau}$ is random: the probability is taken over the draw of $X$. Specifically, it means that if the procedure is repeated on independent samples, at least a fraction $1-\gamma$ of the resulting lower bounds $\underline{\tau}$ lies below $\tau$. In the literature, $1-\gamma$ is called the \emph{nominal level} and $\gamma$ alone the \emph{significance level}.

We define below the notion of \emph{coverage} and \emph{miscoverage}.

\begin{definition}[Coverage \& Miscoverage]
\label{def:coverage}
Let $\underline{\tau}$ be a lower bound on $\tau$, reported with confidence level $1-\gamma$. 
\begin{itemize}
    \item The \emph{coverage probability} is the probability $\mathbb{P}(\tau \ge \underline{\tau})$.
    \item The \emph{miscoverage} is the probability $\mathbb{P}(\tau < \underline{\tau})$.
\end{itemize}  
\end{definition}

A bound is \emph{valid} if, for every value of $\tau$, its coverage is at least its nominal level, i.e., its miscoverage is at most $\gamma$. If the miscoverage is higher than $\gamma$, the lower bound procedure is invalid.

However, validity alone is not enough. For example, the trivial bound $\underline{\tau} = -\infty$ is valid but carries no information, so lower bound procedures often look for the largest valid bound, i.e., the largest bound with coverage at least the nominal level $1-\gamma$.

\subsection{Auditing Formalism}

We follow the framework adopted by prior work on Differential Privacy auditing with Membership Inference Attacks~\cite{yeomPrivacyRiskMachine2018,jagielskiAuditingDifferentiallyPrivate2020a,carliniMembershipInferenceAttacks2022}. Specifically, we adopt the formalism developed by Annamalai et al.~\cite{annamalaiHitchhikersGuideEfficient2025}.

Assume $\mathcal{M}$ satisfies a certain DP variant (like $(\varepsilon, \delta)$-DP, $\mu$-GDP, $(\alpha, \varepsilon)$-RDP~\cite{dworkDifferentialPrivacy, dongGaussianDifferentialPrivacy2019, mironovRenyiDifferentialPrivacy2017}), characterized by a privacy parameter\footnote{For $(\varepsilon, \delta)$-DP auditing, the parameter $\delta$ is often fixed, and only $\varepsilon$ is audited.}.
The goal of an auditing framework is to empirically construct a lower bound for this parameter, with confidence $1-\gamma \in (0, 1)$. More specifically, we consider the following abstraction.

\begin{definition}[Empirical DP Auditing, adapted from~\cite{annamalaiHitchhikersGuideEfficient2025}]
\label{def:empirical-audit}
Let $\tau$ denote the privacy parameter of interest, and let $\tilde\tau$ denote its claimed value. An empirical DP auditing algorithm $\mathsf{Aud}$ takes as input the mechanism $\mathcal{M}$, hyperparameters $\Lambda$, and confidence level $1-\gamma$. It outputs
\[
\underline{\tau} \leftarrow \mathsf{Aud}(\mathcal{M}; \Lambda, 1-\gamma),
\]
a lower bound on $\tau$ with confidence $1-\gamma$.
\end{definition}

The auditing procedure can be used directly to estimate a lower bound of the privacy parameter, but also to verify a claimed guarantee: if $\underline{\tau} > \tilde\tau$, then $\mathcal{M}$ violates its claimed guarantee $\tilde\tau$ with confidence $1-\gamma$.

To illustrate the proposed methodology step-by-step, we introduce the following running example, which will be revisited throughout subsequent sections.

\begin{leftbar}{$\mu$-GDP}
    Assume we have a mechanism $\mathcal{M}$, which satisfies $\mu_0$-GDP. We run an auditing algorithm $\mathsf{Aud}$ with confidence $1 - \gamma$, which outputs $\underline{\mu}$, i.e., $\underline{\mu} \leftarrow \mathsf{Aud}(\mathcal{M}; \emptyset, 1 - \gamma)$. We then expect
    \begin{equation}
        \mathbb{P}(\mu_0 \geq \underline{\mu}) \ge 1 - \gamma.
    \end{equation}
    Here, the probability is taken solely over the randomness of $\mathcal{M}$.
\end{leftbar}

Following prior work on auditing~\cite{jagielskiAuditingDifferentiallyPrivate2020a, nasrTightAuditingDifferentially2023}, we consider a fixed dataset $D \in \mathcal{D}$ and a ``canary'' point $x\in \mathcal{X}$. We construct two datasets $D_0 = D$ and $D_1 = D \cup \{x\}$, which are adjacent under the add/remove relation. Under the DP hypothesis, we have that the distributions $\mathcal{M}(D_0)$ and $\mathcal{M}(D_1)$ are indistinguishable to a certain extent, characterized by the DP guarantee considered and by the value of the parameter.

We define a scoring function $s: \mathcal{W} \times \mathcal{D} \times \mathcal{X} \to \mathbb{R}$. By the post-processing property of DP, the distributions of $s(\mathcal{M}(D_0), D, x)$ and $s(\mathcal{M}(D_1), D, x)$ are at least as indistinguishable as the raw mechanism outputs. This formulation accommodates both black-box and white-box auditing by allowing the scoring function to use the full mechanism output, the dataset, and the canary point.

\subsection{Auditing as a Hypothesis Test}
A membership inference attack is a statistical hypothesis test designed to determine whether the mechanism $\mathcal{M}$ was applied to $D_0$ or $D_1$. Given an observed output $\Theta \in \mathcal{W}$, we define the null and alternative hypotheses:
\begin{align}
    H_0: \Theta &\sim \mathcal{M}(D_0) \\
    H_1: \Theta &\sim \mathcal{M}(D_1)
\end{align}

We evaluate the class of threshold-based membership inference attacks $\mathcal{A}$ of the form $\mathcal{A}^\eta(\Theta, D, x) = \mathds{1}_{\{s(\Theta, D, x) \geq \eta\}}$, where $\Theta \in \mathcal{W}$ denotes the realization of the output of $\mathcal{M}$, $D$ the fixed dataset, 
$x$ the point and $\eta$ is the score threshold, which we can calibrate. The objective of the attack is to distinguish whether the mechanism $\mathcal{M}$ was applied to $D_0$ or $D_1$, i.e., to predict if $x$ was a point of the training set or not. Following standard conventions, a higher score indicates a greater confidence that the training set was $D_1$. 
Consequently, for a given threshold $\eta$, an attack output of $\mathcal{A}^\eta=1$ indicates a positive prediction of membership for $x$.

\begin{leftbar}{DP-SGD}
    If the mechanism $\mathcal{M}$ is the DP-SGD algorithm~\cite{abadiDeepLearningDifferential2016}, its output realization corresponds to the sequence of model parameters $\Theta = (\theta_1, \dots, \theta_T) \in \mathcal{W}$ computed during training. To execute a standard loss-based attack\footnotemark, we would define the scoring function as $s(\Theta, D, x) = - \ell(\theta_T, x)$. 
\end{leftbar}
\footnotetext{The standard loss-based attack ignores the dataset, and only uses the last iterate of the training dynamic. It is a black-box attack.}
We measure the attack's performance at a given threshold $\eta$ using its Type I and Type II error rates. These correspond to the false positive rate (FPR) $\alpha^\eta$ and the false negative rate (FNR) $\beta^\eta$, defined as:
\begin{align}
\label{eq:alpha_beta}
\alpha^\eta &= \mathbb{E}_{H_0}[\mathcal{A}^\eta(\Theta, D, x)] = \mathbb{P}_{H_0}[s(\Theta, D, x) \geq \eta] \\
\beta^\eta &= 1 - \mathbb{E}_{H_1}[\mathcal{A}^\eta(\Theta, D, x)] = \mathbb{P}_{H_1}[s(\Theta, D, x) < \eta]
\end{align}

Any differential privacy guarantee imposes a theoretical limit on the success of such an attack, mathematically bounding the achievable trade-off between  $\alpha^\eta$ and $\beta^\eta$. 
The core principle of empirical auditing is to estimate these error rates using concrete attacks, and use them to derive a robust lower bound for the mechanism's privacy parameter.
\begin{leftbar}{$\mu$-GDP error bounds}
    Under $\mu_0$-GDP~\cite{dongGaussianDifferentialPrivacy2019}, distinguishing $H_0$ from $H_1$ is theoretically bounded by the difficulty of distinguishing $\mathcal{N}(0, 1)$ from $\mathcal{N}(\mu_0, 1)$. For any $\eta$, the attack's error rates must satisfy the following trade-off~\cite{jagielskiAuditingDifferentiallyPrivate2020a}:
    \begin{equation}
    \label{eq:bound_gdp}
        \beta^\eta \geq \Phi\left(\Phi^{-1}(1-\alpha^\eta) - \mu_0\right)
    \end{equation}
    where $\Phi$ is the standard Gaussian CDF. 
\end{leftbar}

To estimate the true probabilities $\alpha^\eta$ and $\beta^\eta$, the natural approach is to run the mechanism $\mathcal{M}$ independently many times, for each hypothesis. For an algorithm such as DP-SGD, this corresponds to training many models on the same dataset, with the specific point $x$ or without, each with a different seed for the injected noise. 
In the language of hypothesis testing, the auditor estimates the error rates $\alpha^\eta$ and $\beta^\eta$, defined as expectations in~\eqref{eq:alpha_beta}, by their Monte-Carlo averages over these runs. Because it requires many executions of DP-SGD, this auditing scheme is commonly referred to as the multi-run protocol.

\subsection{Multi-Run Protocol}
\label{subsec:binomial}
To empirically estimate $\alpha^\eta$ and $\beta^\eta$ in a multi-run protocol, practitioners draw independent samples from the mechanism $\mathcal{M}$ under datasets $D_0$ and $D_1$ to simulate each hypothesis. Specifically, they execute $\mathcal{M}$ independently $n_0$ times to obtain output sequences $\Theta_1^{(0)}, \dots, \Theta_{n_0}^{(0)} \sim \mathcal{M}(D_0)$ and compute the corresponding scores $s_i^{(0)} = s(\Theta_i^{(0)}, D, x)$. Similarly, they run $\mathcal{M}$ independently $n_1$ times on $D_1$ to obtain $\Theta_1^{(1)}, \dots, \Theta_{n_1}^{(1)} \sim \mathcal{M}(D_1)$, computing the scores $s_j^{(1)} = s(\Theta_j^{(1)}, D, x)$.

For a threshold $\eta \in \mathbb{R}$, let $\mathcal{A}_i^{(0), \eta} = \mathds{1}_{\{s_i^{(0)} \geq \eta\}}$ and $\mathcal{A}_j^{(1), \eta} = \mathds{1}_{\{s_j^{(1)} \geq \eta\}}$, denote the binary predictions of the membership inference attacks. Because each mechanism output is generated independently, the variables $\mathcal{A}_1^{(0), \eta}, \dots, \mathcal{A}_{n_0}^{(0), \eta}$ are i.i.d. Bernoulli random variables with parameter $\alpha^\eta$. Similarly, each $1-\mathcal{A}_j^{(1), \eta}$ follows an i.i.d. Bernoulli law of parameter $\beta^\eta$. Their respective sums follow Binomial distributions:
\begin{align}
    \sum_{i=1}^{n_0} \mathcal{A}_i^{(0), \eta} &\sim \operatorname{Binomial}(n_0, \alpha^\eta) \\
    \sum_{j=1}^{n_1} \left(1- \mathcal{A}_j^{(1), \eta}\right) &\sim \operatorname{Binomial}(n_1, \beta^\eta)
\end{align}

Each sum represents the total number of Type I and Type II errors made by the attack. For ease of notation, we define $\hat{k}_0^\eta = \sum_{i=1}^{n_0} \mathcal{A}_i^{(0), \eta}$ and $\hat{k}_1^\eta = \sum_{j=1}^{n_1} (1-\mathcal{A}_j^{(1), \eta})$. We also define the empirical error rates, $\hat{\alpha}^\eta = \hat{k}_0^\eta / n_0$ and $\hat{\beta}^\eta = \hat{k}_1^\eta / n_1$.
Using these error-rate estimates, we seek to recover the privacy parameter.

\subsection{Privacy Parameter Estimation}
To recover the privacy parameter, given empirical error rates, auditing methods invert the trade-off equations to obtain the largest lower bound consistent with these error rates.

To do so, they rely on upper bounding the true error rates $\alpha^\eta, \beta^\eta$ with a certain confidence. Building these upper bounds correctly requires the use of precise statistical tools, and is described in Section~\ref{sec:conf_bound}.

Suppose we have bounded the true error rates $\alpha^\eta$ and $\beta^\eta$ with confidence $1-\gamma/2$ each. 
Specifically, we derived upper bounds $\overline{\alpha}^\eta$ and $\overline{\beta}^\eta$ from their respective empirical estimates $\hat{\alpha}^\eta$ and $\hat{\beta}^\eta$, such that:
\begin{equation}
    \mathbb{P}(\alpha^\eta \le \overline{\alpha}^\eta) \ge 1-\gamma/2 \quad \text{and} \quad
    \mathbb{P}(\beta^\eta \le \overline{\beta}^\eta) \ge 1-\gamma/2.
\end{equation}
These upper bounds can be used to derive a \emph{lower bound} on the privacy parameter, since smaller parameters in DP frameworks denote stronger privacy. Guaranteeing a global $1-\gamma$ confidence for the privacy parameter is achieved by applying a union bound over the two error-rate types, explaining the allocation of a confidence of $1-\gamma/2$ to each.

Doing so involves inverting the DP trade-off condition to obtain a lower bound on the privacy loss parameter, which is the largest one justified by the upper confidence bounds on the error rates. Thus, with confidence $1-\gamma$, the audit establishes that the mechanism's privacy loss parameter is at least this large.
However, the algorithm may be strictly less private, as a more powerful attack could expose even greater privacy leakage, with lower error rates and a stronger lower bound.

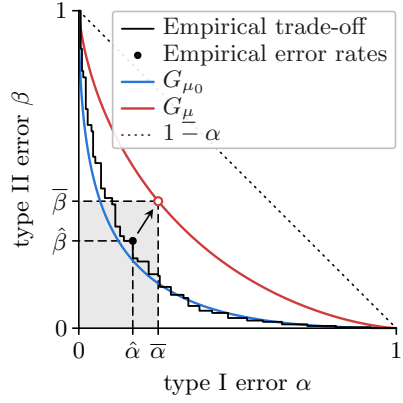
\begin{figure}[t]
    \centering
    \begin{tikzpicture}[
    x=4.2cm, y=4.2cm,
    font=\footnotesize,
    line cap=round, line join=round,
    declare function={
      PhiPoly(\u) = \u*(0.31938153 + \u*(-0.356563782 + \u*(1.781477937
                    + \u*(-1.821255978 + \u*1.330274429))));
      PhiPos(\z)  = 1 - 0.39894228*exp(-0.5*\z*\z)*PhiPoly(1/(1 + 0.2316419*\z));
      Phi(\z)     = 0.5 + sign(\z)*(PhiPos(abs(\z)) - 0.5);
    },
  ]

  \def\muZero{1.75}
      \def\aHat{0.170}    \def\bHat{0.275}
  \def\aBar{0.25}  \def\bBar{0.40}
  \def\muLow{0.927}

  \definecolor{gdpTrue}{HTML}{2a78d6}
  \definecolor{gdpAudit}{HTML}{d03b3b}
  \definecolor{grayBox}{rgb}{0.914, 0.914, 0.914}

  \tikzset{
    tradeoff/.style={line width=0.9pt, domain=-3.8:#1+3.8, samples=90, smooth},
    guide/.style={black, line width=0.6pt, densely dashed},
  }

  \begin{scope}
    \clip (0,0) rectangle (1,1);

    \fill[grayBox] (0,0) rectangle (\aBar,\bBar);

    \draw[black, line width=0.6pt, dotted] (0,1) -- (1,0);

    \draw[guide] (\aHat,\bHat) -- (\aHat,0)  (\aHat,\bHat) -- (0,\bHat);
    \draw[guide] (\aBar,\bBar) -- (\aBar,0)  (\aBar,\bBar) -- (0,\bBar);

    \draw[gdpTrue,  tradeoff=\muZero] plot ({1 - Phi(\x)}, {Phi(\x - \muZero)});
    \draw[gdpAudit, tradeoff=\muLow]  plot ({1 - Phi(\x)}, {Phi(\x - \muLow)});
  \end{scope}

  \tikzset{empiricalStep/.style={draw=black, line width=0.7pt},}
  \draw[empiricalStep]
      (0.000, 1.000) -|
      (0.006, 0.880) -|
      (0.010, 0.810) -|
      (0.014, 0.790) -| 
      (0.022, 0.690) -| 
      (0.028, 0.640) -|
      (0.040, 0.620) -| 
      (0.045, 0.540) -| 
      (0.052, 0.520) -|
      (0.070, 0.440) -|
      (0.082, 0.410) -|
      (0.105, 0.390) -| 
      (0.115, 0.320) -| 
      (0.130, 0.290) -|
      (0.142, 0.275) -|
      (0.170, 0.220) -|
      (0.185, 0.210) -|
      (0.220, 0.170) -|
      (0.250, 0.165) -| 
      (0.255, 0.130) -|
      (0.290, 0.105) -|
      (0.330, 0.095) -| 
      (0.345, 0.070) -|
      (0.380, 0.055) -|
      (0.420, 0.048) -|
      (0.470, 0.032) -|
      (0.530, 0.024) -| 
      (0.580, 0.020) -|
      (0.640, 0.012) -|
      (0.720, 0.007) -|
      (0.810, 0.003) -|
      (0.920, 0.000) -|
      (1.000, 0.000);

  \draw[line width=0.6pt, line cap=rect] (0,1) -- (0,0) -- (1,0);
  \foreach \pos/\lbl in {0/0, 1/1, \aHat/\hat{\alpha}, \aBar/\overline{\alpha}}
    \draw[line width=0.6pt, line cap=butt] (\pos,0) -- ++(0,-2.5pt)
      node[anchor=north, yshift=-1pt, inner sep=0pt, text height=1.6ex, text depth=0pt]
      {$\lbl$};
  \foreach \pos/\lbl in {0/0, 1/1, \bHat/\hat{\beta}, \bBar/\overline{\beta}}
    \draw[line width=0.6pt, line cap=butt] (0,\pos) -- ++(-2.5pt,0)
      node[anchor=east, xshift=-1.5pt, inner sep=0pt] {$\lbl$};
  \node[anchor=north, inner sep=0pt] at ([yshift=-17pt]0.5,0) {type I error $\alpha$};
  \node[anchor=south, inner sep=0pt, rotate=90] at ([xshift=-17pt]0,0.5)
    {type II error $\beta$};

  \draw[black, line width=0.6pt, -stealth, shorten <=3.5pt, shorten >=3.5pt]
    (\aHat,\bHat) -- (\aBar,\bBar);
  \fill (\aHat,\bHat) circle[radius=1.5pt];
  \draw[gdpAudit, fill=white, line width=0.8pt] (\aBar,\bBar) circle[radius=1.5pt];

  \matrix[anchor=north east, inner sep=3pt, row sep=1.5pt, column sep=3pt, rounded corners=1pt,
          draw=black!25, line width=0.6pt, fill=white, fill opacity=0.9,
          nodes={anchor=base west, inner sep=0pt, text height=5.5pt, text depth=3pt}]
    at (1,1) {
    \draw[black, line width=0.7] (0, 2pt) -- (12pt, 2pt);
      & \node{Empirical trade-off}; \\
    \draw[black, fill=black, line width=0.8pt] (6pt, 2pt) circle[radius=1pt];
    & \node{Empirical error rates} ;  \\
    \draw[gdpTrue, line width=0.9pt] (0,2pt) -- (12pt,2pt);
      & \node {$G_{\mu_0}$}; \\
    \draw[gdpAudit, line width=0.9pt] (0,2pt) -- (12pt,2pt);
      & \node {$G_{\underline{\mu}}$}; \\
    \draw[black, line width=0.5pt, dotted] (0,2pt) -- (12pt,2pt);
      & \node[text depth=0pt] {$1-\alpha$}; \\ 
  };

\end{tikzpicture}
    \caption{$\mu$-GDP auditing in the $(\alpha,\beta)$ plane: empirical error rates $(\hat\alpha,\hat\beta)$ of the MIA, their upper confidence bounds $(\overline\alpha,\overline\beta)$, and the trade-off curves of the true $\mu_0$ and of the recovered $\underline\mu$. 
    The gray area corresponds to the valid interval for the error rates. The black point represents the empirical estimate of the attack error rates, at a given threshold.}
    \label{fig:gdp_audit}
\end{figure}

\begin{leftbar}{$\mu$-GDP lower bound}
Given error-rate upper bounds $\overline{\alpha}^\eta$ and $\overline{\beta}^\eta$ that both hold with confidence $1-\gamma/2$ we can construct a valid lower bound $\underline{\mu}^\eta$ for $\mu_0$, with confidence $1-\gamma$, by applying:
\begin{equation}
    \label{eq:mu_lb}
    \underline{\mu}^\eta = \max\left\{\Phi^{-1}(1-\overline{\alpha}^\eta) - \Phi^{-1}(\overline{\beta}^\eta), 0 \right\}.
\end{equation}
The procedure is illustrated in Figure~\ref{fig:gdp_audit}. 
\end{leftbar}

Once we derive a lower bound for the privacy parameter, any mathematically valid transformation of this bound preserves the original confidence level. 
For instance, when evaluating a mechanism such as DP-SGD under the $(\varepsilon, \delta)$-DP framework, we can perform the audit entirely within the $\mu$-GDP framework and derive a lower bound on $(\varepsilon, \delta)$ only at the end, with the same confidence, using Equation~\eqref{eq:eps_to_gdp}. This approach yields tighter empirical bounds~\cite{nasrTightAuditingDifferentially2023} and is therefore preferred over computing $(\varepsilon, \delta)$ bounds directly with $\underline{\varepsilon} = \max \left\{ \ln\left(\frac{1-\overline{\alpha} - \delta}{\overline{\beta}}\right), \ln\left(\frac{1-\overline{\beta} - \delta}{\overline{\alpha}}\right), 0\right\}$.

In the following section, we describe the existing methods to construct upper bounds for the error rates.

\section{Computing Lower Confidence Bounds}
\label{sec:conf_bound}
To build a lower bound on the privacy parameter with confidence $1-\gamma$, we require confidence upper bounds on the estimators $\hat{\alpha}^\eta$ and $\hat{\beta}^\eta$. Such bounds can be obtained using any Binomial proportion confidence interval. We start by describing the methods used in previous audit papers.
\subsection{Prior Methods}
In most auditing implementations, Clopper-Pearson (CP)~\cite{clopperPearson} is chosen as it is an exact method, i.e., it computes bounds from the binomial distribution rather than relying on normal approximations.

\subsubsection{Fixed threshold}
The first valid solution is to fix the threshold $\eta$ in advance (or similarly the FPR level $\alpha$)\footnote{Fixing the FPR level $\alpha$ means fixing the ratio $k/n_0$ of null scores above the threshold in advance and setting $\eta$ to match this ratio. Then $\hat{k}_0^\eta$ is no longer binomial, but $\alpha^\eta \sim \operatorname{Beta}(k+1, n_0-k)$, so the same Clopper-Pearson bound holds.} and to compute the Clopper-Pearson one-sided upper bound for each error type.

\begin{proposition}[Clopper-Pearson one-sided upper bound~\cite{clopperPearson}]
Let $X \sim \operatorname{Bin}(n, p)$ and let $x$ be the observed value. Then with confidence at least $1-\gamma$:

\begin{equation}
    p \le \operatorname{Beta}^{-1}(1 - \gamma; \, x + 1, \, n - x)
\end{equation}
\end{proposition}

We define $\overline{\alpha}^\eta_{1-\gamma/2}:= \operatorname{Beta}^{-1}(1 - \gamma/2; \, \hat{k}_0^\eta + 1, \, n_0 - \hat{k}_0^\eta)$ an upper bound on the true error rate $\alpha^\eta$ with confidence $1-\gamma/2$, such that:
\begin{equation}
    \mathbb{P}(\alpha^\eta \le \overline{\alpha}^\eta_{1-\gamma/2}) \ge 1-\gamma/2
\end{equation}

This method is statistically valid but requires fixing the threshold in advance, which can lead to suboptimal attacks and thus to suboptimal lower bounds for the privacy parameter. The true trade-off curve of the underlying mechanism is unknown, and it can take a different shape than the theoretical boundary defined by the DP framework. Consequently, a blindly fixed threshold may select an FPR where the gap between the mechanism curve and the DP boundary is large, resulting in a loose privacy lower bound.
For example, on the DP-SGD mechanism, prior work~\cite{nasrTightAuditingDifferentially2023} showed that threshold selection has a much larger impact on $(\varepsilon, \delta)$-DP estimation than on $\mu$-GDP estimation. This motivates the need to develop methods to select an appropriate threshold.

\subsubsection{Holdout}
\label{subsec:holdout}
A statistically valid approach is to use a fraction of the data (holdout set) to estimate the optimal threshold, e.g., the threshold that would maximize the value of the privacy parameter lower bound. Then, compute the lower bound on the rest of the data with the attack calibrated with this threshold, using the Clopper-Pearson interval. This method works well in practice and was used in some privacy auditing work~\cite{jagielskiAuditingDifferentiallyPrivate2020a, ganevTightAuditingDifferential2026, tramerDebuggingDifferentialPrivacy2022}. However, it comes with two drawbacks: not all scores are used in the estimation, and the threshold estimate is itself subject to intrinsic noise.

To circumvent this, one idea is to select the optimal threshold on the same set of scores.

\subsubsection{Optimal threshold}
\label{subsec:max}

To be able to select this optimal threshold $\eta$ on the same set of scores, we would like to obtain a bound that holds \emph{simultaneously} across multiple thresholds. However, applying the standard Clopper-Pearson interval directly with confidence $1-\gamma/2$ over each error type, and then selecting the threshold that yields the best lower bound is invalid. It creates a \emph{selection bias}.

\begin{leftbar}{$\mu$-GDP selection bias}
    Let $\underline{\mu}^\eta_{1-\gamma}$ denote the lower bound for the $\mu_0$-GDP mechanism, derived with confidence 
    $1-\gamma$, using the upper bounds $\overline{\alpha}^\eta_{1-\gamma/2}, \overline{\beta}^\eta_{1-\gamma/2}$, and Equation~\eqref{eq:mu_lb}. In the privacy auditing literature, it is common practice to report the maximum value found across the set of all possible thresholds $\{\eta_k\}_k$,
    \begin{equation}
        \underline{\mu}^\mathrm{\max} := \max_{\eta_k} \underline{\mu}_{1-\gamma}^{\eta_k}
    \end{equation}
     as the overall lower bound. This lower bound is invalid for the desired confidence level $1-\gamma$.
\end{leftbar}

 However, we observe that this method is commonly used in the privacy auditing literature, yielding lower bounds that fail to guarantee the claimed confidence level~\cite{nasrTightAuditingDifferentially2023, steinkePrivacyAuditingOne2023, mahloujifarAuditingfDifferentialPrivacy2024, annamalaiNearlyTightBlackBox2024, cebereTighterPrivacyAuditing2025, maddockCANIFECraftingCanaries2023}.

The confidence interval guarantees that at least a fraction $1-\gamma$ of the time, the estimated lower bound does not exceed the true parameter. By selecting the optimal threshold without corrections, these guarantees no longer hold. In practice, it means that the claimed empirical lower bound can exceed the true parameter more than a fraction $\gamma$ of the time. This matters because the two errors are not symmetric. A loose lower bound is still valid, merely less informative. An inflated one could produce false alarms during an audit of a correct implementation.

Although Nasr et al.~\cite{nasrTightAuditingDifferentially2023} demonstrated that every point on the theoretical $\mu$-GDP trade-off curve yields the same lower bound, we emphasize that this invariance holds only for the \emph{population} distribution. Because the \emph{empirical} distribution contains sampling noise, selecting the threshold that yields the maximum lower bound and applying a classical Clopper-Pearson confidence interval overfits to this noise, violating the lower bound's statistical guarantees. This theoretical argument has been used in multiple papers~\cite{nasrTightAuditingDifferentially2023, annamalaiNearlyTightBlackBox2024, cebereTighterPrivacyAuditing2025}, sometimes incorrectly justifying the use of the optimal threshold selected on the same set, without corrections.

We then propose three additional statistically valid methods, adapted from the multiple-comparison literature.

\subsection{Union Bound and Bonferroni Correction}
To still be able to use the optimal threshold on the same set, one straightforward solution is to apply a union bound and to divide each threshold's confidence parameter proportionately. In the context of simultaneous statistical inference, this is known as the Bonferroni correction~\cite{bonferroniCorrection}.

\begin{proposition}[Union Bound, Bonferroni Correction~\cite{bonferroniCorrection}]
Let $E_1, \dots, E_m$ be events on a common probability space such that $\mathbb{P}(E_k)\ge 1-\gamma_k$, for each $k \in \{1,\dots,m\}$. Then,
\begin{equation}
    \mathbb{P}\left(\bigcap_{k=1}^m E_k\right) \ge 1 - \sum_{k=1}^m \gamma_k.
\end{equation}
In particular, setting $\gamma_k = \gamma/m$ for all $k$ gives $\mathbb{P}\left(\bigcap_{k=1}^m E_k\right) \ge 1-\gamma$.
\end{proposition}

In practice, to apply this correction, we must fix a finite grid of thresholds $\eta_1, \dots, \eta_m$ independently of the data. For each $k$, let  $\overline{\alpha}^{\eta_k}_{1-\gamma/2m}$ be the upper bound defined by Clopper-Pearson with confidence $1-\gamma/2m$, for the observed count $\hat{k}_0^{\eta_k}$. Then the event $E_k = (\alpha^{\eta_k} \leq \overline{\alpha}^{\eta_k}_{1-\gamma/2m})$ holds with probability at least $1-\gamma/2m$. The Bonferroni correction guarantees with confidence at least $1-\gamma/2$, simultaneously for every $k \in \{1,\dots,m\}$:
\begin{equation}
    \alpha^{\eta_k} \le \overline{\alpha}^{\eta_k}_{1-\gamma/2m}.
\end{equation}

Because this holds jointly across the grid, the auditor may select a threshold \emph{after} observing the data, e.g., the $\eta_k$ yielding the best empirical trade-off, and the guarantee still applies to whichever $\eta_k$ is chosen, with no further correction needed.

The correction factor $\gamma/2m$ is the statistical penalty for committing to a finite set of thresholds in advance.
Concretely, this makes each individual bound $\overline{\alpha}^{\eta_k}_{1-\gamma/2m}$ looser, with the looseness growing as the grid size $m$ increases.
The next result removes this restriction by bounding $\alpha^\eta$ uniformly over $\eta \in \mathbb{R}$, so no grid needs to be pre-specified.

\subsection{DKW Inequality}

\begin{proposition}[Dvoretzky-Kiefer-Wolfowitz One-Sided Inequality~\cite{dkw, dkw_massart}]
Let $X_1, \dots, X_n$ be i.i.d.\ real-valued random variables with CDF $F$, and let $\hat{F}_n(t) = \frac{1}{n}\sum_{i=1}^n \mathds{1}_{\{X_i \le t\}}$ be the empirical CDF. Then for any $\gamma \in (0,1/2)$:
\begin{equation}
    \mathbb{P}\left(\sup_{t \in \mathbb{R}} \left(\hat{F}_n(t) - F(t)\right) > \sqrt{\frac{\ln(1/\gamma)}{2n}}\right) \le \gamma
\end{equation}
\end{proposition}
For the auditing framework, let $F_0$ be the CDF of the score $s(\mathcal{M}(D_0), D, x)$ under $H_0$, and $\hat{F}_0$ its empirical counterpart, estimated from the $n_0$ samples $s_1^{(0)}, \dots, s_{n_0}^{(0)}$. We have $\alpha^\eta = 1 - F_0(\eta^-)$ and $\hat{\alpha}^\eta = 1-\hat F_0(\eta^-)$, where the superscript $\eta^-$ denotes the left-hand limit of the cumulative distribution function at $\eta$. The DKW inequality on $\hat F_0$ transfers to the empirical error rates $\hat{\alpha}^\eta, \hat{\beta}^\eta $, giving, with confidence at least $1-\gamma/2$, simultaneously for every $\eta \in \mathbb{R}$:
\begin{equation}
\label{eq:dkw_bound}
    \alpha^\eta \le \hat{\alpha}^\eta + \sqrt{\frac{\ln(2/\gamma)}{2n_0}},
\end{equation}
and similarly for $\beta^\eta$ with $n_1$ instead of $n_0$.

\begin{figure}[tbp]
    \centering
    \adjustbox{margin=0pt 0pt 12mm 0pt}{\includegraphics{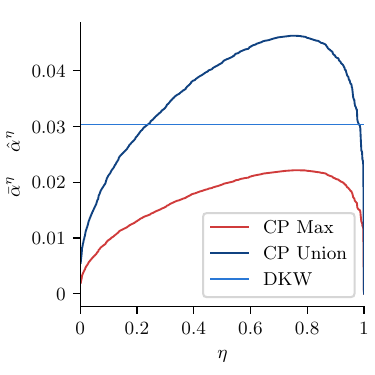}}
    \caption{Confidence interval size $\overline{\alpha}^\eta - \hat{\alpha}^\eta$ depending on the threshold $\eta$ for different confidence methods at a fixed confidence level $1-\gamma = 0.95$.}
    \label{fig:conf_size}
\end{figure}

DKW holds with a finite number of samples, and is distribution-free. It is not related to the binomial property of our estimators. Consequently, it is conservative for any fixed $F_0$ and might result in looser bounds. As shown by Equation~\eqref{eq:dkw_bound}, the confidence interval size, $\overline{\alpha}^\eta - \hat{\alpha}^\eta$, is independent of $\eta$, leading to unusable bounds in the very low or very high FPR regions. This phenomenon is illustrated in Figure~\ref{fig:conf_size} where we compare the size of the confidence intervals, when varying the threshold $\eta$ for the different methods presented. As expected, the confidence interval is narrower when selecting the max threshold than when using the union bound.

\subsection{\texorpdfstring{$K$-Fold}{K-Fold}}
\label{subsec:kfold}
The $K$-fold selection, borrowed from cross-validation, uses a mix of different previous methods. The observation scores are split into $K$ folds, the threshold is selected on one holdout fold, and the bound is computed on the remaining $K-1$ folds. This yields $K$ valid lower bounds, each computed on $1-1/K$ of the data. We report the largest one, which remains valid if each fold is tested at level $\gamma/K$ (union bound over the folds). We use $K=5$ in all our experiments. To our knowledge, this method has not been used in prior privacy auditing work.

\section{One-Run Methods}
\label{sec:one_run}

While natural, the multi-run protocol might be impractical as it requires training numerous models, which is computationally costly. To decrease this cost, Steinke et al.~\cite{steinkePrivacyAuditingOne2023} developed one-run methods which extract the required scores from a single training by inserting many canaries at once. In the following section, we demonstrate how this one-run protocol exhibits the same selection bias, and we discuss which of the tools developed for multi-run confidence intervals remain applicable.

\subsection{One-Run Protocol}

This one-run protocol is made possible by randomizing, independently, the training inclusion of each canary, and then bounding the attacker's guessing accuracy without requiring the per-canary scores to be independent.

Specifically, the auditor selects $m$ target points and assigns each an independent membership bit $b_k \sim \operatorname{Bernoulli}(0.5)$ for $k \in \{1, \dots, m\}$. The mechanism is executed exactly once to produce a single model $\Theta \sim \mathcal{M}(D \cup \{x_k \mid b_k = 1\})$. The attacker then computes the scores $s_k = s(\Theta, D, x_k)$.
The framework allows choosing two thresholds $\eta_{+} > \eta_{-}$, instead of one, and setting the predictions:
\begin{equation}
    \mathcal{A}_k^\eta = \left\{\begin{array}{cl}
        1 & \text{if}\quad s_k \ge \eta_{+} \\
        0 & \text{if} \quad s_k \le \eta_{-} \\
        \bot & \text{if} \quad  \eta_{-} < s_k < \eta_{+}
    \end{array}\right.
\end{equation}
for all $m$ canaries simultaneously. The $\bot$ means the score is discarded, and the prediction is not used as a guess. In the following, for ease of notation, $\eta$ will represent the pair $(\eta_{-}, \eta_{+})$. In practice, the thresholds are set by ranks: the attack guesses ``in'' for the $k_+$ largest scores and ``out'' for the $k_-$ smallest ones, so that the number of guesses $r^\eta = k_+ + k_-$ is fixed in advance.

Because all canaries are evaluated against the same model $\Theta$, their scores $s_k$ are correlated. The empirical error counts, $\hat k^\eta_{0}, \hat k^\eta_{1}$, defined in Section~\ref{subsec:binomial}, no longer follow independent Binomial distributions. The Clopper-Pearson upper bound cannot be applied directly to obtain a valid confidence interval once $\eta$ has been selected.

Despite this correlation, the theoretical framework abstracts the attack into a generic guessing game and derives a bound that holds regardless of how the $s_k$ are correlated given the $b_k$. The core test statistic is the total number of correct guesses, $c^\eta = \sum_{\mathcal{A}_k^\eta \neq \bot} \mathds{1}_{\{\mathcal{A}^\eta_k = b_k\}}$.
Under $(\varepsilon, \delta)$-DP, Steinke et al.~\cite{steinkePrivacyAuditingOne2023} derive a theoretical upper bound on the probability of achieving at least $c^\eta$ correct guesses given a fixed $r^\eta$ guesses.
Their result is presented below.

\begin{proposition}[One Run DP bound~\cite{steinkePrivacyAuditingOne2023}, informal]
    \label{prop:one-run-bound}
    If $\mathcal{M}$ is $(\varepsilon, \delta)$-DP, for all $v \in \mathbb{R}$: \[
        \mathbb{P}\left( c^\eta \geq v\right) \leq g_{\varepsilon, \delta}(m, v, r^\eta)
    \]
    for an explicit function $g$ depending only on $m, v, r^\eta$,  and $(\varepsilon, \delta)$. This result holds irrespective of any correlation among the $\mathcal{A}_k$'s induced by their shared dependence on $\Theta$.
\end{proposition}
Informally, this proposition states that, given an $(\varepsilon, \delta)$-DP mechanism and $m$ injected canaries, the probability of making at least $v$ correct guesses out of $r^\eta$ guesses is bounded. This result is the mathematical equivalent of the bounds on the error rates for independent scores, as presented for $\mu$-GDP in Equation~\eqref{eq:bound_gdp}.

This probability is then used to provide a lower bound on the privacy parameter, with a confidence $1-\gamma$.

\begin{equation}
\label{eq:one_run_lb}
    \underline{\varepsilon}^\eta_{1-\gamma} = \sup\left\{\varepsilon \geq 0 : g_{\varepsilon, \delta}(m, c^\eta, r^\eta) < \gamma \right\}.
\end{equation}

While first derived for the $(\varepsilon, \delta)$-DP framework, it has been adapted by Mahloujifar et al.~\cite{mahloujifarAuditingfDifferentialPrivacy2024} to match the $f$-DP framework~\cite{dongGaussianDifferentialPrivacy2019}, by keeping the same protocol but changing the way the probability is computed.

\subsection{Selection bias}
\label{subsec:one_run_selection}

The lower bound~\eqref{eq:one_run_lb} is valid for a number of guesses fixed before observing the scores. As in the multi-run protocol, prior work selects it afterwards. Both Steinke et al.~\cite{steinkePrivacyAuditingOne2023} and Mahloujifar et al.~\cite{mahloujifarAuditingfDifferentialPrivacy2024} compute the lower bound for several numbers of guesses\footnote{The arXiv version of~\cite{steinkePrivacyAuditingOne2023} mentions that this can be resolved by lowering the confidence, though this detail is omitted in the published conference version.}, i.e., thresholds $\{\eta_j\}_j$, and report the largest value,
\begin{equation}
\label{eq:one_run_max}
    \underline{\varepsilon}^{\max} := \max_{\eta_j} \underline{\varepsilon}^{\eta_j}_{1-\gamma}.
\end{equation}

For the same reason as for the multi-run method, this procedure invalidates the bound: the confidence level is not guaranteed.

\subsection{Methods}
\label{subsec:one_run_ci}
Most of the corrections proposed in Section~\ref{sec:conf_bound} apply to the one-run protocol, with the number of guesses in place of the threshold. 
Fixing $(k_-, k_+)$ in advance is valid, and provides a correct lower bound.
The union bound method applied to a grid of numbers of guesses fixed in advance, and with a corrected confidence level for each bound, is also valid.
The same holds for the holdout method, which selects the number of guesses on a random half of the canaries and computes the bound from the guesses on the other half. This also remains valid.
The $K$-fold selection follows in the same way, with the bound computed on the $K-1$ remaining folds at level $\gamma / K$. 

However, the DKW inequality \emph{does not apply}: it requires independent samples, and no bound holds uniformly over the number of guesses.

\section{Experiments}
\label{sec:experiments}

To illustrate our theoretical results, and quantify the impact of this selection bias, we ran experiments, both on synthetic distributions with perfect scores, and on DP models trained on real-world datasets.

The code and scripts required to reproduce all experiments will be made publicly available upon acceptance.

\begin{figure*}[htbp]
    \centering
    \includegraphics[width=\textwidth]{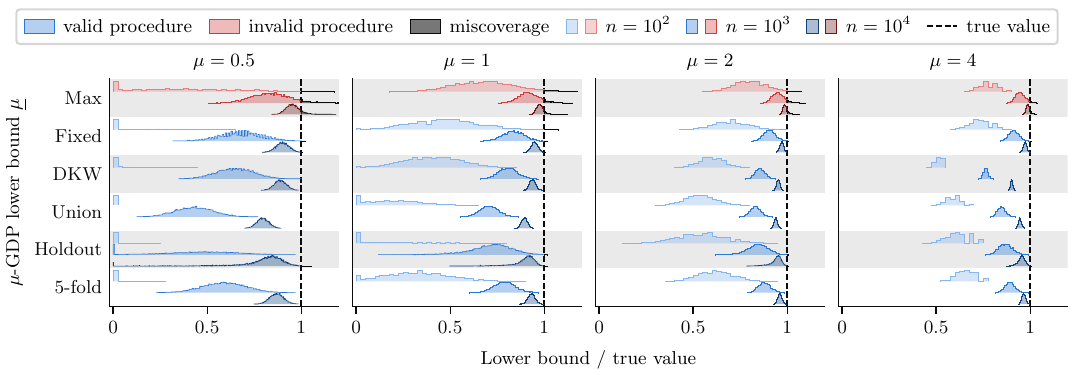}
    \caption{Distribution of the lower bound $\underline{\mu}$ reported by each procedure, as a fraction of the true parameter with confidence $1-\gamma = 95\%$. The underlying mechanism is $\mu$-GDP and the attack considered is the optimal test. Each histogram represents 10,000 audits, with darker shading reflecting a larger number of samples. The part in dark gray, beyond the true value, corresponds to the audits whose lower bound exceeds the true parameter (miscoverage).}
    \label{fig:distribution_mu_gdp_95}
\end{figure*}

\subsection{Miscoverage of Lower Bounds Methods}
We first tested our framework over generated distributions to measure the empirical coverage of each method presented in Section~\ref{sec:conf_bound}, and whether there are settings in which the lower bound is overestimated. As defined in Definition~\ref{def:coverage}, the coverage of a statistical lower bound corresponds to the probability that the lower bound is lower than the parameter we seek to estimate. It can be estimated by running multiple independent audits, and computing the fraction of time in which the lower bound lies below the true parameter. 
The same holds for the miscoverage: the \emph{empirical miscoverage} is the fraction of audits in which the lower bound exceeds the true parameter. For a valid method, it should not exceed the significance level $\gamma$.

Computing it requires knowing the true parameter. We therefore simulate a Gaussian mechanism, for which the scores are drawn from $\mathcal{N}(0, 1)$ under $H_0$ and from $\mathcal{N}(\mu, 1)$ under $H_1$. Thresholding these scores is the optimal test, so the theoretical trade-off curve of the attack is exactly $G_{\mu}$, and the true parameter is $\mu$. For each $\mu \in \{0.5, 1, 2, 4\}$ and each number of samples per audit $n_0 = n_1 = n \in \{10^2, 10^3, 10^4\}$, we run $10\,000$ independent audits, and compute the $\mu$-GDP lower bound of each method at confidence $1-\gamma = 95\%$. As all the samples were generated independently, this experiment corresponds to the multi-run protocol. Table~\ref{tab:T1_coverage_95} reports the resulting empirical miscoverage.

The invalid selection (Max) exceeds the nominal level $\gamma = 5\%$ in all settings with $n \ge 10^3$ except one, and its miscoverage increases with $n$, up to $19.4\%$ for $\mu_0 = 2$ and $n = 10^4$, almost four times the nominal level. In contrast, as expected, all the valid methods stay below the nominal level. Figure~\ref{fig:distribution_mu_gdp_95} illustrates this by showing the histograms of the computed lower bound $\underline{\mu}$, over the different methods presented.

\begin{table}[htbp]
    \centering
    \caption{Empirical miscoverage (\%) of the $\mu$-GDP lower bound at confidence $1-\gamma = 95\%$ on a Gaussian mechanism with an optimal attack.  $10\,000$ audits per cell, $n$ samples per hypothesis. \textbf{Bold:} above significance level $\gamma=5\%$.}
    \setlength{\tabcolsep}{4pt}
    \label{tab:T1_coverage_95}
    \small
    \begin{tabular}{llrrrrrr}
\toprule
\multicolumn{2}{c}{Parameters} &\multicolumn{6}{c}{Empirical Miscoverage (\%)} \\
\midrule
$\mu$ & $n$ & Max (Invalid) & Fixed & Union & DKW & Holdout & $K$-fold \\

\midrule
0.5 & 100 & 3.29 & 0.17 & 0.00 & 0.04 & 0.12 & 0.04 \\
 & 1000 & \textbf{8.23} & 0.24 & 0.00 & 0.06 & 0.20 & 0.07 \\
 & 10000 & \textbf{13.44} & 0.29 & 0.00 & 0.07 & 0.37 & 0.06 \\
\addlinespace[4pt]
1 & 100 & 2.94 & 0.25 & 0.00 & 0.01 & 0.14 & 0.03 \\
 & 1000 & \textbf{9.01} & 0.26 & 0.00 & 0.02 & 0.26 & 0.04 \\
 & 10000 & \textbf{16.42} & 0.18 & 0.00 & 0.02 & 0.26 & 0.06 \\
\addlinespace[4pt]
2 & 100 & 1.79 & 0.13 & 0.00 & 0.00 & 0.04 & 0.02 \\
 & 1000 & \textbf{8.89} & 0.25 & 0.00 & 0.00 & 0.24 & 0.07 \\
 & 10000 & \textbf{19.36} & 0.24 & 0.00 & 0.00 & 0.42 & 0.11 \\
\addlinespace[4pt]
4 & 100 & 0.00 & 0.00 & 0.00 & 0.00 & 0.00 & 0.00 \\
 & 1000 & 3.44 & 0.23 & 0.00 & 0.00 & 0.12 & 0.04 \\
 & 10000 & \textbf{15.34} & 0.15 & 0.00 & 0.00 & 0.32 & 0.05 \\
\bottomrule
\end{tabular}
\end{table}

\subsection{Comparison of Methods}
\begin{figure}[htbp]
    \centering
    \adjustbox{margin=0pt 0pt 6mm 0pt}{\includegraphics{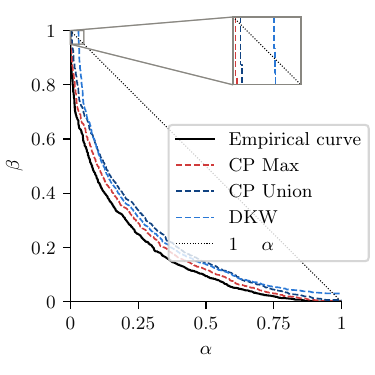}}
    \caption{Empirical trade-off curve (plain line) and computed upper bounds for different methods (dashed lines) plotted in the $(\alpha, \beta)$ plane. The underlying mechanism is $\mu$-GDP with $\mu=2$, and upper bounds are computed with $n_0 = n_1=2000$, and $1-\gamma=0.95$. The gray inset zooms in on the low FPR regime $\alpha \in [0, 0.05]$.}
    \label{fig:upper_bounds_comparison}
\end{figure}

Using the same model, we also compare the trade-off curve produced by the upper bounds on $\alpha^\eta$ and $\beta^\eta$ of the three methods that provide a bound at every threshold: CP Max, CP Union and DKW. 
For each threshold $\eta$, we place the point $(\overline{\alpha}^\eta, \overline{\beta}^\eta)$ in the $(\alpha, \beta)$ plane, creating an upper-bound estimate for the trade-off curve.
Figure~\ref{fig:upper_bounds_comparison} presents the different trade-off curve upper bounds as well as the empirical trade-off curve. As expected, the upper bound found using the invalid max method lies below the other upper bound curves.

This figure also illustrates the limit of the constant-size interval produced by DKW: in the very low or very high FPR regions, its upper bounds become greater than the random response trade-off curve $1-\alpha$, which makes them unusable. This can be an issue when evaluating the $(\varepsilon, \delta)$-DP framework, where tight audits of DP-SGD are often only computable in the very low or very high FPR regions~\cite{nasrTightAuditingDifferentially2023, carliniMembershipInferenceAttacks2022}. 

\subsection{Recomputing Prior Bounds}
We selected some auditing works from the literature that mostly use the methods described in Section~\ref{subsec:max}, and recomputed the claimed bounds, with the other presented valid methods. Table~\ref{tab:prior_audits} summarizes the works selected, and the method they were using to compute the lower bound.

For each work, we selected their main experiment and recomputed the bounds using all the different valid methods, and the one presented in the original paper. To be able to compare our results with the ones presented in the different papers, we only changed the way the lower-bound on the privacy parameter was computed, not how the rest of the audit was performed. For example, it means that for articles that did not use the latest auditing methods leveraging $\mu$-GDP~\cite{nasrTightAuditingDifferentially2023}, we kept their original audit setup and only updated the bound calculation.

We present the results grouped by type of audit: multi-run, one-run and tight black-box audits. For each group, we only report the main results here: the detailed experimental protocols are given in Appendix~\ref{app:exp_details}, and the results of all the methods in Appendix~\ref{app:add_results}, Table~\ref{tab:results}.

\begin{table}[h]
    \centering

    \caption{Selected prior works, with the method they used to select the threshold, and the corresponding statistical validity.}
    \label{tab:prior_audits}
    \begin{threeparttable}
        \setlength{\tabcolsep}{4pt}
        \small
        \begin{tabular}{lll}
\toprule
Paper & Threshold selection & Statistical validity \\
\midrule
Jagielski et al.~\cite{jagielskiAuditingDifferentiallyPrivate2020a} & Holdout & Valid \\
Nasr et al.~\cite{nasrTightAuditingDifferentially2023} & Max & Selection bias$^{*}$ \\
Steinke et al.~\cite{steinkePrivacyAuditingOne2023} & Max & Selection bias$^{*}$ \\
Mahloujifar et al.~\cite{mahloujifarAuditingfDifferentialPrivacy2024} & Max & Selection bias \\
Annamalai \& De Cristofaro~\cite{annamalaiNearlyTightBlackBox2024} & Max & Selection bias$^{*}$ \\
Cebere et al.~\cite{cebereTighterPrivacyAuditing2025} & Max & Selection bias\\
\bottomrule
\addlinespace[2pt]
\multicolumn{3}{@{}l@{}}{\small $^{*}$ Acknowledged by the authors in the paper.} \\

\end{tabular}
    \end{threeparttable}
\end{table}

\subsubsection{Multi-run auditing}
We selected two influential auditing works that do their audit following the multi-run protocol: the mechanism is run many times with and without a canary, each run yields one score, and the error rates of the attack that thresholds these scores are bounded with the Clopper-Pearson interval.

The first one performs a black-box audit using a ClipBKD procedure to find the canary point~\cite{jagielskiAuditingDifferentiallyPrivate2020a}. This paper uses a valid procedure to find the threshold (holdout\footnote{They use a variant of the holdout method we presented, which explains the slightly different results in Table~\ref{tab:results}. 
Instead of selecting the threshold that would maximize the Clopper-Pearson bound itself (our method), they select the one that maximizes the ratio of the empirical true positive rate to the empirical false positive rate, $(1-\hat{\beta}^\eta)/\hat{\alpha}^\eta$.}), but it can be improved, using the union bound or a $K$-fold. 
We selected the experiment that trains a two-layer neural network for classification on Fashion-MNIST~\cite{xiao2017fashionmnistnovelimagedataset} and reproduced it for $\varepsilon_{\text{th}} \in \{1, 2, 4, 8\}$,  with $1-\gamma=0.99$. We find similar values to those presented in their paper, with their threshold selection method. By using a $K$-fold for selecting the threshold, with $K=5$, instead of a holdout set, we were able to improve their bounds by up to $40\%$. 

In~\cite{nasrTightAuditingDifferentially2023}, the authors introduced an efficient tight white-box protocol, and the use of $\mu$-GDP to efficiently convert the $\overline{\alpha}, \overline{\beta}$ bounds to $\underline{\varepsilon}$. Nevertheless, they select the threshold that yields the maximum final lower bound without any correction. Among the different experiments they report, we selected the experiment which provides the tightest bounds. We use their white-box protocol with the DP training of a CNN~\cite{cnn} on CIFAR-10~\cite{cifar10}, to compute a lower bound $\underline{\varepsilon}$ with confidence $1-\gamma=0.95$. For low values of $\varepsilon_\text{th}$ in particular, the valid methods yield much lower bounds than their selection of the optimal threshold on the same scores, meaning that a large part of the reported lower bound is not supported at the claimed confidence level.

For these two multi-run papers, the results for the two best methods (DKW and $K$-fold) are presented in Table~\ref{tab:results_multirun}. The full results are detailed in Table~\ref{tab:results} in Appendix~\ref{app:add_results}.

\begin{table}[htbp] 
    \centering
    
    \caption{Computation of lower bounds $\underline{\varepsilon}$ for prior multi-run audits using different methods. Each result is averaged over 16 seeds,  and the $\pm$ represents 
    $1$ standard deviation. Best valid results are in \textbf{bold}. Diff ($\Delta\%$): relative difference, in \%, between the best valid bound and the bound of the paper's method.}
        
    \small
    \definecolor{lightgray}{rgb}{0.914, 0.914, 0.914}
    \begin{threeparttable}
        
        \setlength{\tabcolsep}{4pt}
        \begin{tabular}{@{} p{3cm} c c c c c c @{}}
\toprule
  & Theoretical & Paper's method & DKW & $K$-fold & Diff \\
 \cmidrule(lr){2-2} \cmidrule(lr){3-3} \cmidrule(lr){3-5} \cmidrule(l){6-6}
Paper & $\varepsilon_\text{th}$ & $\underline\varepsilon$ & $\underline\varepsilon$ & $\underline\varepsilon$ & $\Delta\%$ \\
\midrule
\multirow{4}{=}{Jagielski et al.~\cite{jagielskiAuditingDifferentiallyPrivate2020a}} & $1$ & $0.15_{\pm0.03}$ & $0.17_{\pm0.02}$ & $\mathbf{0.20}_{\pm0.04}$ & $+40\%$ \\
  & $2$ & $0.37_{\pm0.04}$ & $0.37_{\pm0.01}$ & $\mathbf{0.52}_{\pm0.05}$ & $+39\%$ \\
  & $4$ & $0.69_{\pm0.07}$ & $0.59_{\pm0.01}$ & $\mathbf{0.88}_{\pm0.06}$ & $+27\%$ \\
  & $8$ & $0.94_{\pm0.06}$ & $0.70_{\pm0.00}$ & $\mathbf{1.10}_{\pm0.05}$ & $+16\%$ \\
\addlinespace[4pt]
\multirow{4}{=}{Nasr et al.~\cite{nasrTightAuditingDifferentially2023}} & $1$ & $0.54_{\pm0.62}\cellcolor{lightgray}$ & $0.01_{\pm0.02}$ & $\mathbf{0.04}_{\pm0.15}$ & $-93\%$ \\
 & $4$ & $3.11_{\pm0.85}\cellcolor{lightgray}$ & $\mathbf{1.84}_{\pm0.68}$ & $1.43_{\pm0.75}$ & $-41\%$ \\
 & $8$ & $6.83_{\pm1.03}\cellcolor{lightgray}$ & $\mathbf{5.38}_{\pm0.77}$ & $4.89_{\pm0.79}$ & $-21\%$ \\
 & $16$ & $14.08_{\pm1.75}\cellcolor{lightgray}$ & $\mathbf{11.95}_{\pm1.18}$  & $11.39_{\pm1.25}$ & $-15\%$ \\
\bottomrule
\end{tabular}
\begin{tablenotes}[flushleft]
\small
\item \textcolor{lightgray}{\rule[0ex]{3ex}{1.5ex}} Number of guesses chosen on the same scores without correction: not a valid lower bound at confidence \(1-\gamma\).
\end{tablenotes}

    \end{threeparttable}
    \label{tab:results_multirun}
\end{table}

\subsubsection{One-run auditing}
We selected the two main works~\cite{steinkePrivacyAuditingOne2023, mahloujifarAuditingfDifferentialPrivacy2024} that do their audit following the one-run protocol described in Section~\ref{sec:one_run}. We consider their white-box audits, as one-run audits are not yet tight in the black-box setting.

Steinke et al.~\cite{steinkePrivacyAuditingOne2023} introduced the one-run protocol, with a lower bound computed in the $(\varepsilon, \delta)$ framework. They compute this bound for several numbers of guesses and report the largest bound. We selected their white-box experiment that trains a WRN-16~\cite{wrn} on CIFAR-10~\cite{cifar10} with $m=5\,000$ canaries. We reproduced it for $\varepsilon_{\text{th}} \in \{1, 2, 4, 8\}$, using the same values as them. For each method, we computed a lower bound $\underline{\varepsilon}$ with confidence $1-\gamma=0.95$. The best valid method yields a lower bound that is $34\%$ smaller at $\varepsilon_\text{th}=1$ and $13\%$ smaller at $\varepsilon_\text{th} = 8$.

We then reproduced the results of Mahloujifar et al.~\cite{mahloujifarAuditingfDifferentialPrivacy2024} by using the exact same protocol, but tightening the bound using the $f$-DP framework. Like them, we used the $\mu$-GDP framework to estimate a lower bound $\underline{\mu}$ before converting it back to a lower bound $\underline{\varepsilon}$ using Equation~\eqref{eq:eps_to_gdp}. The correction has a similar effect, from $-33\%$ at $\varepsilon_\text{th} = 1$ to $-15\%$ at $\varepsilon_\text{th} = 8$, and their bound remains tighter than the one of Steinke et al.\ once corrected.

The results for the two best methods on this set of experiments (holdout and $K$-fold) are presented in Table~\ref{tab:results_onerun}.

\begin{table}[htbp] 
    \centering
    \caption{Computation of lower bounds $\underline{\varepsilon}$ for prior one-run audits using different methods. Each result is averaged over 16 seeds,  and the $\pm$ represents 
    $1$ standard deviation. Best valid results are in \textbf{bold}. Diff ($\Delta\%$): relative difference, in \%, between the best valid bound and the bound of the paper's method.}
    \small
    \label{tab:results_onerun}
    \begin{threeparttable}
        \setlength{\tabcolsep}{4pt}
        \definecolor{lightgray}{rgb}{0.914, 0.914, 0.914}
        \begin{tabular}{@{} p{3cm} c c c c c c @{}}
 \toprule
  & Theoretical & Paper's method & Holdout & $K$-fold & Diff \\
 \cmidrule(lr){2-2} \cmidrule(lr){3-5} \cmidrule(r){6-6}
Paper & $\varepsilon_\text{th}$ & $\underline\varepsilon$ & $\underline\varepsilon$ & $\underline\varepsilon$ & $\Delta\%$ \\
\midrule
\multirow{4}{=}{Steinke et al.~\cite{steinkePrivacyAuditingOne2023}} & $1$ & $0.55_{\pm0.16}\cellcolor{lightgray}$ & $0.34_{\pm0.18}$ & $\mathbf{0.36}_{\pm0.09}$ & $-34\%$ \\
  & $2$ & $1.13_{\pm0.28}\cellcolor{lightgray}$ & $0.77_{\pm0.36}$ & $\mathbf{0.85}_{\pm0.13}$ & $-24\%$ \\
  & $4$ & $2.04_{\pm0.20}\cellcolor{lightgray}$ & $1.65_{\pm0.20}$ & $\mathbf{1.67}_{\pm0.12}$ & $-18\%$ \\
  & $8$ & $3.57_{\pm0.27}\cellcolor{lightgray}$ & $\mathbf{3.09}_{\pm0.33}$ & $3.05_{\pm0.17}$ & $-13\%$ \\
\addlinespace[4pt]
\multirow{4}{=}{Mahloujifar et al.~\cite{mahloujifarAuditingfDifferentialPrivacy2024}} & $1$ & $0.70_{\pm0.17}\cellcolor{lightgray}$ & $0.41_{\pm0.20}$ & $\mathbf{0.47}_{\pm0.11}$ & $-33\%$ \\
  & $2$ & $1.44_{\pm0.31}\cellcolor{lightgray}$ & $0.98_{\pm0.39}$ & $\mathbf{1.06}_{\pm0.14}$ & $-26\%$ \\
  & $4$ & $2.69_{\pm0.22}\cellcolor{lightgray}$ & $\mathbf{2.13}_{\pm0.28}$ & $2.10_{\pm0.15}$ & $-21\%$ \\
  & $8$ & $5.18_{\pm0.42}\cellcolor{lightgray}$ & $\mathbf{4.39}_{\pm0.60}$ & $4.10_{\pm0.31}$ & $-15\%$ \\
\bottomrule
\end{tabular}
\begin{tablenotes}[flushleft]
\small
\item \textcolor{lightgray}{\rule[0ex]{3ex}{1.5ex}} Number of guesses chosen on the same scores without correction: not a valid lower bound at confidence \(1-\gamma\).
\end{tablenotes}

    \end{threeparttable}
    
\end{table}

\subsubsection{Tight black-box auditing}
Finally, we selected two recent works that seek tight audits when the adversary only observes the final model~\cite{annamalaiNearlyTightBlackBox2024, cebereTighterPrivacyAuditing2025}. They both follow the multi-run protocol, and use the $\mu$-GDP conversion~\cite{nasrTightAuditingDifferentially2023} to have a tight $\underline{\varepsilon}$ bound. They both select the threshold that yields the maximum lower bound on the same scores, without any correction.

Annamalai \& De Cristofaro~\cite{annamalaiNearlyTightBlackBox2024} obtain nearly tight black-box audits by training a CIFAR-10~\cite{cifar10} model from a worst-case initialization. We selected their last-layer fine-tuning experiment and reproduced it for $\varepsilon_\text{th} \in \{1, 2, 4, 10\}$, with $1-\gamma=0.95$ and $500$ models per hypothesis. The best valid method yields a lower bound that is $74\%$ smaller at $\varepsilon_\text{th} = 1$ and $12\%$ smaller at $\varepsilon_\text{th} = 10$.

Cebere et al.~\cite{cebereTighterPrivacyAuditing2025} audit DP-SGD in the hidden state threat model, with an adversary that has access only to the last iterate of the model and its initialization.
We selected their experiment on the Housing dataset~\cite{housing}, where they simulate a specific training trajectory by injecting canary gradients (they call it $\mathcal{A}_\text{GC}$-R).
We reproduced it for $\varepsilon_\text{th} \in \{1, 4, 8, 12, 24\}$, with injected canary gradients at every step. We compute all the scores with $1-\gamma=0.95$ and consider $5\,000$ runs per audit. The correction has a smaller effect than for most other audits, from $-32\%$ at $\varepsilon_\text{th} = 1$ to $-5\%$ at $\varepsilon_\text{th} = 24$.

For these two experiments, the results are presented in Table~\ref{tab:results_tight}, for the methods giving the best lower bounds (DKW and $K$-fold here).

\begin{table}[htbp] 
    \centering
    
    \caption{Computation of lower bounds $\underline{\varepsilon}$ for prior tight black-box audits using different methods. Each result is averaged over 16 seeds,  and the $\pm$ represents 
    $1$ standard deviation. Best valid results are in \textbf{bold}. Diff ($\Delta\%$): relative difference, in \%, between the best valid bound and the bound of the paper's method.}
    \label{tab:results_tight}

    \small
    \begin{threeparttable}
        \setlength{\tabcolsep}{4pt}
        \definecolor{lightgray}{rgb}{0.914, 0.914, 0.914}
        \begin{tabular}{@{} p{3cm} c c c c c c @{}}

\toprule
  & Theoretical & Paper's method & DKW & $K$-fold & Diff \\
 \cmidrule(lr){2-2} \cmidrule(lr){3-5} \cmidrule(r){6-6}
Paper & $\varepsilon_\text{th}$ & $\underline\varepsilon$ & $\underline\varepsilon$ & $\underline\varepsilon$ & $\Delta\%$ \\
\midrule
\addlinespace[4pt]
\multirow{4}{=}{Annamalai \& De Cristofaro~\cite{annamalaiNearlyTightBlackBox2024}} & $1$ & $0.63_{\pm0.24}\cellcolor{lightgray}$ & $\mathbf{0.16}_{\pm0.18}$ & $0.09_{\pm0.13}$ & $-74\%$ \\
 & $2$ & $1.52_{\pm0.24}\cellcolor{lightgray}$ &  $\mathbf{1.01}_{\pm0.28}$ & $0.84_{\pm0.31}$ & $-34\%$ \\
 & $4$ & $3.47_{\pm0.38}\cellcolor{lightgray}$  & $\mathbf{2.74}_{\pm0.32}$ & $2.73_{\pm0.38}$ & $-21\%$ \\
 & $10$ & $9.14_{\pm0.46}\cellcolor{lightgray}$ & $7.70_{\pm0.30}$ & $\mathbf{8.06}_{\pm0.32}$ & $-12\%$ \\
\addlinespace[6pt]

\multirow{5}{=}{Cebere et al.~\cite{cebereTighterPrivacyAuditing2025}} & $1$ & $0.72_{\pm0.18}\cellcolor{lightgray}$ & $\mathbf{0.49}_{\pm0.17}$ & $0.40_{\pm0.18}$ & $-32\%$ \\
 & $4$ & $3.31_{\pm0.45}\cellcolor{lightgray}$ & $\mathbf{3.01}_{\pm0.40}$ & $2.99_{\pm0.40}$ & $-9\%$ \\
 & $8$ & $7.11_{\pm0.91}\cellcolor{lightgray}$ & $6.56_{\pm0.84}$ & $\mathbf{6.60}_{\pm0.85}$ & $-7\%$ \\
 & $12$ & $10.57_{\pm1.44}\cellcolor{lightgray}$ & $9.57_{\pm1.24}$ & $\mathbf{9.86}_{\pm1.46}$ & $-7\%$ \\
 & $24$ & $20.02_{\pm2.89}\cellcolor{lightgray}$  & $16.92_{\pm1.91}$ & $\mathbf{18.94}_{\pm2.72}$ & $-5\%$ \\
\bottomrule
\end{tabular}
\begin{tablenotes}[flushleft]
\small
\item \textcolor{lightgray}{\rule[0ex]{3ex}{1.5ex}} Number of guesses chosen on the same scores without correction: not a valid lower bound at confidence \(1-\gamma\).
\end{tablenotes}
    \end{threeparttable}

\end{table}

\section{Discussion and Limitations}
In the following section, we discuss our results, their implications for prior and future audits as well as the limitations of our methods. 

\subsection{Discussion}

Our results show that addressing the selection bias is a requirement to obtain meaningful lower bounds, not a minor methodological adjustment. On the Gaussian mechanism in Table~\ref{tab:T1_coverage_95}, the miscoverage reaches $19.4\%$, almost four times what should be permitted by the significance level $\gamma = 5\%$. On models trained on real datasets, replacing the bound obtained by the invalid optimal threshold by the bound of the best valid method, on the same set of scores, lowers the value of the lower bound by $5\%$ to $93\%$ (Tables~\ref{tab:results_multirun},~\ref{tab:results_onerun} and~\ref{tab:results_tight}). These differences are important and cannot be neglected when trying to provide a reliable audit.

Nevertheless, as detailed below in Section~\ref{subsec:limitations}, part of this gap comes from the conservativeness of the valid methods and not from the bias alone. Unlike in our idealized Gaussian setting, these audits are not tight, so the true parameter is unknown. Therefore, computing the empirical coverage, by running the same experiments dozens of times, would only show that the computed lower bounds do not exceed the theoretical values for the specific current attack. With a stronger one, the same method could exceed the guarantee more often than the significance level allows. Validity is, in fact, a property of the method, not of one experiment. 
Our Gaussian experiment, where the attack is optimal, suggests that both effects contribute: the miscoverage of the max method exceeds the significance level, while those of the valid methods stay far below it.

We demonstrated that this issue arises across multiple auditing methods. We observed it for the multi-run and the one-run protocols, in tight and non-tight settings, in the white-box or black-box settings. It also holds for every privacy level we tested, but not with the same gap. As the model becomes less private, the relative difference (in $\%$) between the invalid method and the best valid method decreases. Increasing the number of samples does not seem to solve the issue. As presented in Figure~\ref{fig:distribution_mu_gdp_95}, on the Gaussian mechanism, while the max bounds get closer to the valid bound when the number of samples increases, the miscoverage worsens.

Finally, no valid method dominates. DKW and $K$-fold give the best bounds for the multi-run audits, holdout and $K$-fold for the one-run audits, where DKW does not apply. $K$-fold is the only method that is the best or the second best in every experiment, and its variance is lower than that of holdout in most settings, as it does not rely on a single split to find the threshold. 

\subsection{Implications}

The lower bounds reported by the works that select the threshold, or the number of guesses, on the same scores, without any correction, do not hold at the confidence level they might claim. However, they should not be entirely discarded; they remain informative estimates of the privacy leakage. 
But they should not be used as certified lower bounds. 
For example, in our experiments, for $\varepsilon_\text{th} \geq 4$, the bounds of the white-box audit of Nasr et al.~\cite{nasrTightAuditingDifferentially2023} reach $78\%$ to $88\%$ of the theoretical guarantee $\varepsilon_{\text{th}}$ when computed with the max method, but only $46\%$ to $75\%$ once corrected. Using them as certified lower bounds would lead one to conclude that the privacy analysis is almost tight, while a valid bound leaves a gap more than twice as large.

When an audit is used to verify an implementation, or the compliance with a privacy level imposed by a regulator, it needs a certified lower bound, not an empirical estimate. In this context, a lower bound above the claimed guarantee would be read as a violation~\cite{nasrTightAuditingDifferentially2023, tramerDebuggingDifferentialPrivacy2022}.  With an invalid selection of the threshold, a correct implementation could be flagged as faulty in up to one audit out of five, instead of the one out of twenty announced by the confidence level. The use of a valid confidence method to produce the bound removes these false alarms, but comes at a cost: detecting the actual violations requires a tighter audit.

For future audits, we recommend selecting the threshold, or the number of guesses, with a valid method. This method should be chosen before observing the scores, as reporting the best of several valid bounds is again a selection. We also encourage researchers and future auditors to report carefully how this threshold was selected. Overall, without knowledge of the best method for a specific problem, $K$-fold seems to be a reasonable default.

\subsection{Limitations}
\label{subsec:limitations}

The valid methods we presented are all conservative. As presented in Table~\ref{tab:T1_coverage_95}, they all have a miscoverage that sits far below the significance level $\gamma$, which means they do not fully ``exploit'' it. It indicates that tighter valid bounds exist, and new methods could be developed to find them.

The miscoverage can even drop below the significance value for invalid methods (e.g., for $n=100$ in Table~\ref{tab:T1_coverage_95}). It means that for this specific experiment the bound could be valid even with the optimal threshold selected on the same dataset. However, verifying it would require repeating the experiment dozens of times to estimate the coverage probability, which is often too costly for practical applications. This is the reason why, for the experiments on real models, we do not compute the coverage. We could only measure how much the bounds were decreasing when applying valid methods.

Finally, we only consider frequentist confidence bounds, and leave Bayesian credible intervals~\cite{zanella-beguelinBayesianEstimationDifferential2022} to future work.

\section{Conclusion}
A privacy auditing algorithm reports a lower bound that holds at a given confidence level. This guarantee is critical when the audit is used to verify an implementation or to check compliance with a required privacy level. We show that this guarantee no longer holds when the MIA threshold is selected on the same scores used to compute the bound. Yet, selecting the optimal threshold on the same set of scores is common in the auditing literature.

We present multiple valid alternatives, and compute their empirical miscoverage against the invalid procedure for a Gaussian mechanism. To quantify the impact of this selection, we recompute the lower bounds of prior practical audits with the valid alternatives, across several protocols and privacy levels, in both the white-box and black-box settings.

Our results do not question the attacks or the auditing frameworks developed in prior work, but only the confidence at which their bounds hold. For future audits, we recommend that practitioners choose a valid selection method before observing the scores, and report it clearly. As all the valid methods we studied are conservative, a natural direction for future work is to develop tighter lower-bound methods.

\section*{LLM usage considerations}

In this manuscript, LLMs were used for rephrasing sentences only, and all outputs were checked by the authors. LLMs were not used for any mathematical content, structure of the paper or design of experiments. LLMs were used as coding assistants to draw figures or parallelize some experiments. The authors reviewed and tested any code produced by LLMs.

\section*{Acknowledgements}

This work is part of the project SSF-ML-DH of the PEPR Digital Health and was supported by the French National Research Agency, as a part of the France 2030 program, under grant ANR-22-PESN-0014-01. Furthermore, we thank GENCI at IDRIS for providing computer and storage resources through the grant 2026-AD011017108 on the supercomputer Jean Zay.

We also acknowledge funding from CIFAR through Nicolas Papernot's Canada CIFAR AI Chair. Resources used in preparing this research were provided, in part, by the Province of Ontario, the Government of Canada through CIFAR, and companies sponsoring the Vector Institute. We acknowledge the support of the Inria International Chair.

Finally, we are grateful to Mohammad Yaghini and Emmy Fang for their helpful feedback.

\bibliographystyle{IEEEtran}
\bibliography{preprint}

\appendix

\section{Experimental details}
\label{app:exp_details}
\subsection{\texorpdfstring{Reproducing Jagielski et al.~\cite{jagielskiAuditingDifferentiallyPrivate2020a}}{Reproducing Jagielski et al.}}
\label{app:jagielski}
We re-run the ClipBKD audit of Jagielski et al.~\cite{jagielskiAuditingDifferentiallyPrivate2020a} with their protocol. We chose to re-run the protocol with Fashion-MNIST~\cite{xiao2017fashionmnistnovelimagedataset}, restricted to classes $0$ and $1$ as in their paper. We use their two-layer network  and a fixed initialization. We use DP-SGD with clipping norm of $1$, batches of $250$ for $24$ epochs with a learning rate $0.15$. We do not add any $\ell_2$ regularization. They report bounds at confidence $1-\gamma=0.99$.

We use the RDP accountant for $\delta = 10^{-5}$ and $\varepsilon_{\mathrm{th}} \in \{1, 2, 4, 8\}$. The canary $x$ is their ClipBKD point, inserted $k=4$ times\footnote{When inserting $k$ times the same point, we are working under group-DP~\cite{dworkCalibratingNoiseSensitivity} so we divide the estimated $\hat{\varepsilon}$ by $k$.}. For each $(\varepsilon_{\mathrm{th}}, k)$, $1000$ models are trained without the canary and $1000$ with it. The first $500$ of each are used to select the threshold, and on the other $500$, they compute one-sided Clopper-Pearson upper bounds with $1-\gamma/2$.

\subsection{\texorpdfstring{Reproducing Nasr et al.~\cite{nasrTightAuditingDifferentially2023}}{Reproducing Nasr et al.}}
\label{app:nasr}
We reproduced the results presented in~\cite{nasrTightAuditingDifferentially2023} using their white-box protocol, on the ScatterNet CNN of~\cite{tramerDifferentiallyPrivateLearning2021} trained on CIFAR-10~\cite{cifar10}. We computed lower bounds for $\varepsilon_\text{th} \in \{1, 4, 8, 16\}$ and $\delta = 10^{-5}$ using their best method, $f$-DP (CP).
We used DP-SGD with a learning rate of $4.0$, a clipping constant of $0.1$, a batch size of $4\,096$, and a momentum of $0.9$ for $60$ epochs. We report scores with $1-\gamma=0.95$ confidence. At $\varepsilon=8$,  models have a test accuracy of $\approx 69.5\%$.
For the canaries, we use gradient ``Dirac'' canaries (their clipped gradients equal $C$ at a single coordinate, distinct across canaries, and zero elsewhere) that are inserted at each step.

\subsection{\texorpdfstring{Reproducing Steinke et al.~\cite{steinkePrivacyAuditingOne2023}}{Reproducing Steinke et al.}}
We reproduced their white-box audit on CIFAR-10~\cite{cifar10}. The model is a WRN-16-4~\cite{wrn} in which batch normalization is replaced by group normalization (16 groups), trained with the high-accuracy DP-SGD recipe of De et al.~\cite{de2022unlockinghighaccuracydifferentiallyprivate} as implemented by Sander et al.~\cite{pmlr-v202-sander23b}. They use a batch size of $4\,096$, a clipping norm of $1$, and a plain SGD with a learning rate of $4$, for $T = 2\,500$ steps. They also use an exponential moving average (EMA) of the weights with decay 0.9999.

We trained our models using DP-SGD with $\varepsilon_\text{th} \in \{1, 2, 4, 8\}$, and $\delta=10^{-5}$.
We added $m = 5\,000$ auditing examples to the full training set ($n = 50\,000$), each included independently with probability $1/2$. The auditing examples are gradient ``Dirac'' canaries~\cite{nasrTightAuditingDifferentially2023} and the white-box scores are $\sum_t \langle w^{t-1} - w^t, \hat g_i\rangle$. We guess ``in'' for the $k$ largest scores and ``out'' for the $k$ smallest. The selection bias happens when choosing $k$.

\subsection{\texorpdfstring{Reproducing Mahloujifar et al.~\cite{mahloujifarAuditingfDifferentialPrivacy2024}}{Reproducing Mahloujifar et al.}}
Their CIFAR-10 experiment reuses the setting above: a WRN-16-4 with group normalization trained as in~\cite{de2022unlockinghighaccuracydifferentiallyprivate,pmlr-v202-sander23b} with batch size $4\,096$, and $2\,500$ steps, audited with $m = 5\,000$ Dirac canaries. We therefore audit the very same training runs, canaries and guesses, and only replace the estimator with theirs: a recursive $f$-DP bound on the number of correct guesses, tested against the trade-off curves of the Gaussian mechanism, the smallest rejected noise being converted to $\varepsilon$ at $\delta = 10^{-5}$. As in their paper, the number of guesses is swept and the best value kept, with the same corrections as above.

\subsection{\texorpdfstring{Reproducing Annamalai \& De Cristofaro~\cite{annamalaiNearlyTightBlackBox2024}}{Reproducing Annamalai \& De Cristofaro}}
We reproduced the last-layer fine-tuning audit of~\cite{annamalaiNearlyTightBlackBox2024}: a logistic regression trained with full-batch DP-SGD ($20$ steps, learning rate $4\cdot10^{-4}$, clipping norm $1$, $\delta=10^{-5}$) on the WRN-28-10 features~\cite{de2022unlockinghighaccuracydifferentiallyprivate} of $25\,000$ CIFAR-10~\cite{cifar10} images, with their released worst-case initialization and ClipBKD target~\cite{jagielskiAuditingDifferentiallyPrivate2020a}. For $\varepsilon_\text{th}\in\{1,2,4,10\}$ we train $500$ models with and $500$ without the target, score each by the loss of the target, and apply their estimator: one-sided Clopper-Pearson bounds on the error rates at the threshold maximizing the bound on the same observations, converted to $\mu$-GDP and then to $\varepsilon$ at $1-\gamma=0.95$.

\subsection{\texorpdfstring{Reproducing Cebere et al.~\cite{cebereTighterPrivacyAuditing2025}}{Reproducing Cebere et al.}}
We reproduced the audit of~\cite{cebereTighterPrivacyAuditing2025} in the hidden state threat model, where the adversary observes the final model. In their experiments, the adversary does not insert a data point but a crafted Dirac gradient. We reproduced one of their experiments, the one named $\mathcal{A}_\text{GC}$-R with a periodicity of $k=1$ (meaning the crafted gradient is inserted at every iteration). For the score, they consider the difference between $\theta_T$ and $\theta_0$ on the same coordinate as the Dirac crafted gradient. We train a small fully connected network on the Housing dataset~\cite{housing} for $250$ steps, for $\varepsilon_\text{th} \in \left\{1, 4, 8, 12, 24 \right\}$ and $\delta = 10^{-5}$. Each audit uses $5\,000$ runs, the crafted gradient being inserted with probability $1/2$ in each, and is repeated across $16$ seeds.

\section{Additional Results}
\label{app:add_results}

We report in Table~\ref{tab:T1_coverage_99} the empirical miscoverage of each method on the Gaussian mechanism at confidence $1-\gamma = 99\%$ in the same setting as Table~\ref{tab:T1_coverage_95}. In Table~\ref{tab:results}, we report the lower bounds obtained with every method for all the audits evaluated.

\begin{table}[htbp]
\centering
\caption{Empirical miscoverage (\%) of the $\mu$-GDP lower bound at confidence $1-\gamma = 99\%$ on a Gaussian mechanism with an optimal attack.  $10\,000$ audits per cell, $n$ samples per hypothesis. \textbf{Bold:} above significance level $\gamma=1\%$.}
\label{tab:T1_coverage_99}
\small
\setlength{\tabcolsep}{4pt}
\begin{tabular}{llrrrrrr}
\toprule
\multicolumn{2}{c}{Parameters} &\multicolumn{6}{c}{Empirical Miscoverage (\%)} \\
\midrule
$\mu$ & $n$ & Max (Invalid) & Fixed & Union & DKW & Holdout & $K$-fold \\
\midrule
0.5 & 100 & 0.18 & 0.02 & 0.00 & 0.00 & 0.00 & 0.00 \\
 & 1000 & 0.72 & 0.02 & 0.00 & 0.00 & 0.01 & 0.00 \\
 & 10000 & \textbf{1.19} & 0.02 & 0.00 & 0.00 & 0.01 & 0.01 \\
\addlinespace[2pt]
1 & 100 & 0.16 & 0.01 & 0.00 & 0.00 & 0.00 & 0.00 \\
 & 1000 & 0.81 & 0.01 & 0.00 & 0.00 & 0.02 & 0.00 \\
 & 10000 & \textbf{2.04} & 0.01 & 0.00 & 0.00 & 0.00 & 0.00 \\
\addlinespace[2pt]
2 & 100 & 0.08 & 0.02 & 0.00 & 0.00 & 0.00 & 0.00 \\
 & 1000 & 0.92 & 0.03 & 0.00 & 0.00 & 0.02 & 0.00 \\
 & 10000 & \textbf{2.97} & 0.00 & 0.00 & 0.00 & 0.03 & 0.00 \\
\addlinespace[2pt]
4 & 100 & 0.00 & 0.00 & 0.00 & 0.00 & 0.00 & 0.00 \\
 & 1000 & 0.29 & 0.02 & 0.00 & 0.00 & 0.00 & 0.00 \\
 & 10000 & \textbf{2.51} & 0.00 & 0.00 & 0.00 & 0.02 & 0.01 \\ 
\bottomrule
\end{tabular}

\end{table}

\begin{table*}[htbp]
    \centering
    
    \caption{Computation of lower bounds $\underline{\varepsilon}$ for all prior audits using different methods. $\varepsilon_\text{th}$ is the guarantee of the audited mechanism. $\underline\varepsilon$ is the lower bound at the paper's confidence level $1-\gamma$, with the paper's own estimator (their method) and with the corrected estimators of Section~\ref{sec:conf_bound}. Each result is averaged over 16 seeds. The $\pm$ represents 
    $1$ standard deviation. Best valid results are in \textbf{bold}. Diff ($\Delta\%$) is the relative difference, in \%, between the best valid bound and the bound of the paper's method. Protocols are detailed in Appendix~\ref{app:exp_details}.}
    \footnotesize
    \setlength{\tabcolsep}{5pt}
    \definecolor{lightgray}{rgb}{0.914, 0.914, 0.914}
    \label{tab:results}
    \begin{tabularx}{\textwidth}{@{} >{\raggedright\arraybackslash}X c c c c c c c c @{}}
\toprule
 & Confidence & Theoretical & Paper's method & Union & DKW & Holdout & $K$-fold & Diff. \\
\cmidrule(lr){2-2} \cmidrule(lr){3-3} \cmidrule(lr){4-4} \cmidrule(lr){5-8} \cmidrule(lr){9-9}
Paper & $1-\gamma$ & $\varepsilon_\text{th}$ & $\underline\varepsilon$ & $\underline\varepsilon$ & $\underline\varepsilon$ & $\underline\varepsilon$ & $\underline\varepsilon$ & $\Delta$ \\
\midrule
\multirow{4}{=}{Jagielski et al.~\cite{jagielskiAuditingDifferentiallyPrivate2020a}} & \multirow{4}{*}{$0.99$} & $1$ & $0.15_{\pm0.03}$ & $0.18_{\pm0.03}$ & $0.17_{\pm0.02}$ & $0.16_{\pm0.08}$ & $\mathbf{0.20}_{\pm0.04}$ & $+40\%$ \\
 &  & $2$ & $0.37_{\pm0.04}$ & $0.48_{\pm0.03}$ & $0.37_{\pm0.01}$ & $0.46_{\pm0.08}$ & $\mathbf{0.52}_{\pm0.05}$ & $+39\%$ \\
 &  & $4$ & $0.69_{\pm0.07}$ & $0.83_{\pm0.04}$ & $0.59_{\pm0.01}$ & $0.79_{\pm0.08}$ & $\mathbf{0.88}_{\pm0.06}$ & $+27\%$ \\
 &  & $8$ & $0.94_{\pm0.06}$ & $1.04_{\pm0.01}$ & $0.70_{\pm0.00}$ & $1.03_{\pm0.06}$ & $\mathbf{1.10}_{\pm0.05}$ & $+16\%$ \\
\addlinespace[4pt]
\multirow{4}{=}{Nasr et al.~\cite{nasrTightAuditingDifferentially2023}} & \multirow{4}{*}{$0.95$} & $1$ & $0.54_{\pm0.62}\cellcolor{lightgray}$ & $0.00$ & $0.01_{\pm0.02}$ & $0.00$ & $\mathbf{0.04}_{\pm0.15}$ & $-93\%$ \\
 &  & $4$ & $3.11_{\pm0.85}\cellcolor{lightgray}$ & $0.55_{\pm0.50}$ & $\mathbf{1.84}_{\pm0.68}$ & $0.92_{\pm0.91}$ & $1.43_{\pm0.75}$ & $-41\%$ \\
 &  & $8$ & $6.83_{\pm1.03}\cellcolor{lightgray}$ & $3.62_{\pm0.71}$ & $\mathbf{5.38}_{\pm0.77}$ & $3.66_{\pm1.92}$ & $4.89_{\pm0.79}$ & $-21\%$ \\
 &  & $16$ & $14.08_{\pm1.75}\cellcolor{lightgray}$ & $9.52_{\pm1.03}$ & $\mathbf{11.95}_{\pm1.18}$ & $9.58_{\pm3.10}$ & $11.39_{\pm1.25}$ & $-15\%$ \\
\addlinespace[4pt]
\multirow{4}{=}{Steinke et al.~\cite{steinkePrivacyAuditingOne2023}} & \multirow{4}{*}{$0.95$} & $1$ & $0.55_{\pm0.16}\cellcolor{lightgray}$ & $0.35_{\pm0.10}$ & -- & $0.34_{\pm0.18}$ & $\mathbf{0.36}_{\pm0.09}$ & $-34\%$ \\
 &  & $2$ & $1.13_{\pm0.28}\cellcolor{lightgray}$ & $0.79_{\pm0.08}$ & -- & $0.77_{\pm0.36}$ & $\mathbf{0.85}_{\pm0.13}$ & $-24\%$ \\
 &  & $4$ & $2.04_{\pm0.20}\cellcolor{lightgray}$ & $1.60_{\pm0.09}$ & -- & $1.65_{\pm0.20}$ & $\mathbf{1.67}_{\pm0.12}$ & $-18\%$ \\
 &  & $8$ & $3.57_{\pm0.27}\cellcolor{lightgray}$ & $2.79_{\pm0.12}$ & -- & $\mathbf{3.09}_{\pm0.33}$ & $3.05_{\pm0.17}$ & $-13\%$ \\
\addlinespace[4pt]
\multirow{4}{=}{Mahloujifar et al.~\cite{mahloujifarAuditingfDifferentialPrivacy2024}} & \multirow{4}{*}{$0.95$} & $1$ & $0.70_{\pm0.17}\cellcolor{lightgray}$ & $0.45_{\pm0.11}$ & -- & $0.41_{\pm0.20}$ & $\mathbf{0.47}_{\pm0.11}$ & $-33\%$ \\
 &  & $2$ & $1.44_{\pm0.31}\cellcolor{lightgray}$ & $1.03_{\pm0.18}$ & -- & $0.98_{\pm0.39}$ & $\mathbf{1.06}_{\pm0.14}$ & $-26\%$ \\
 &  & $4$ & $2.69_{\pm0.22}\cellcolor{lightgray}$ & $2.08_{\pm0.15}$ & -- & $\mathbf{2.13}_{\pm0.28}$ & $2.10_{\pm0.15}$ & $-21\%$ \\
 &  & $8$ & $5.18_{\pm0.42}\cellcolor{lightgray}$ & $4.02_{\pm0.30}$ & -- & $\mathbf{4.39}_{\pm0.60}$ & $4.10_{\pm0.31}$ & $-15\%$ \\
\addlinespace[4pt]
\multirow{4}{=}{Annamalai \& De Cristofaro~\cite{annamalaiNearlyTightBlackBox2024}} & \multirow{4}{*}{$0.95$} & $1$ & $0.63_{\pm0.24}\cellcolor{lightgray}$ & $0.00_{\pm0.01}$ & $\mathbf{0.16}_{\pm0.18}$ & $0.04_{\pm0.08}$ & $0.09_{\pm0.13}$ & $-74\%$ \\
 &  & $2$ & $1.52_{\pm0.24}\cellcolor{lightgray}$ & $0.45_{\pm0.23}$ & $\mathbf{1.01}_{\pm0.28}$ & $0.48_{\pm0.38}$ & $0.84_{\pm0.31}$ & $-34\%$ \\
 &  & $4$ & $3.47_{\pm0.38}\cellcolor{lightgray}$ & $2.18_{\pm0.28}$ & $\mathbf{2.74}_{\pm0.32}$ & $2.15_{\pm0.77}$ & $2.73_{\pm0.38}$ & $-21\%$ \\
 &  & $10$ & $9.14_{\pm0.46}\cellcolor{lightgray}$ & $7.36_{\pm0.30}$ & $7.70_{\pm0.30}$ & $7.17_{\pm1.23}$ & $\mathbf{8.06}_{\pm0.32}$ & $-12\%$ \\
\addlinespace[4pt]
\multirow{5}{=}{Cebere et al.~\cite{cebereTighterPrivacyAuditing2025}} & \multirow{5}{*}{$0.95$} & $1$ & $0.72_{\pm0.18}\cellcolor{lightgray}$ & $0.20_{\pm0.14}$ & $\mathbf{0.49}_{\pm0.17}$ & $0.30_{\pm0.26}$ & $0.40_{\pm0.18}$ & $-32\%$ \\
 &  & $4$ & $3.31_{\pm0.45}\cellcolor{lightgray}$ & $2.66_{\pm0.41}$ & $\mathbf{3.01}_{\pm0.40}$ & $2.66_{\pm0.56}$ & $2.99_{\pm0.40}$ & $-9\%$ \\
 &  & $8$ & $7.11_{\pm0.91}\cellcolor{lightgray}$ & $6.23_{\pm0.83}$ & $6.56_{\pm0.84}$ & $6.36_{\pm0.90}$ & $\mathbf{6.60}_{\pm0.85}$ & $-7\%$ \\
 &  & $12$ & $10.57_{\pm1.44}\cellcolor{lightgray}$ & $9.46_{\pm1.34}$ & $9.57_{\pm1.24}$ & $9.41_{\pm1.44}$ & $\mathbf{9.86}_{\pm1.46}$ & $-7\%$ \\
 &  & $24$ & $20.02_{\pm2.89}\cellcolor{lightgray}$ & $18.16_{\pm2.55}$ & $16.92_{\pm1.91}$ & $18.32_{\pm2.92}$ & $\mathbf{18.94}_{\pm2.72}$ & $-5\%$ \\
\bottomrule
\addlinespace[2pt]
\multicolumn{9}{@{}l@{}}{\scriptsize \textcolor{lightgray}{\rule[0ex]{3ex}{1.5ex}} Threshold or number of guesses chosen on the same scores without correction: not a valid lower bound at confidence $1-\gamma$.} \\
\end{tabularx}
\end{table*}

\end{document}